\documentclass[12pt]{article}
\usepackage{graphicx}
\usepackage{a4wide}
\usepackage{epsfig}
\newcommand{\eps}{\varepsilon}
\newcommand{\pfrac}[2]{\left(\frac{#1}{#2}\right)}
\newcommand{\slk}{/\kern-6pt k}
\newcommand{\sll}{/\kern-4pt l}
\newcommand{\slp}{p\kern-5pt/}
\newcommand{\slq}{q\kern-5.5pt/}
\newcommand{\slx}{x\kern-5.5pt/}
\newcommand{\dDk}{\frac{d^Dk}{(2\pi)^D}}
\newcommand{\dDl}{\frac{d^Dl}{(2\pi)^D}}

\newcommand{\bbbone}{\hbox{\rm 1\kern-3pt l}}
\newcommand{\ms}{{\rm MS}}
\newcommand{\msbar}{{\overline{\rm MS}}}
\newcommand{\Tr}{\mathop{\rm Tr}\nolimits}
\newcommand{\GeV}{\mathop{\rm GeV}\nolimits}
\newcommand{\Disc}{\mathop{\rm Disc}\nolimits}
\newcommand{\tfrac}[2]{\textstyle\frac{#1}{#2}}
\newcommand{\Frac}[2]{{\textstyle\frac{#1}{#2}}}
\newcommand{\Ei}{\mathop{\rm Ei}\nolimits}
\newdimen\brackheight
\newdimen\bracklength
\newcommand{\overbrack}[3]{\brackheight=3pt\multiply\brackheight by#1
  \bracklength=1pt\multiply\bracklength by#3
  \raise12pt\hbox to0pt{\kern#2pt\vbox{\hrule height0.3pt\hbox{\vrule
  width0.3pt\hbox{\vbox to\brackheight{\hbox
  to\bracklength{\hfill}\vfill}}\vrule width0.3pt}}\hss}}

\begin{document}

\thispagestyle{empty} 
\begin{flushright}
MITP/14-003
\end{flushright}
\vspace{0.1cm}

\begin{center}
{\Large\bf Perturbative $O(\alpha_s)$ corrections to the correlation\\[7pt] 
functions of light tetraquark currents}\\[1.3cm]
{\large S.~Groote$^{1,2}$, J.G.~K\"orner$^2$ and D.~Niinepuu$^1$}\\[1cm]
$^1$ Loodus- ja Tehnoloogiateaduskond, F\"u\"usika Instituut,\\[.2cm]
  Tartu \"Ulikool, T\"ahe 4, 51010 Tartu, Estonia\\[7pt]
$^2$PRISMA Cluster of Excellence, Institut f\"ur Physik,
  Johannes-Gutenberg-Universit\"at,\\[.2cm]
  Staudinger Weg 7, 55099 Mainz, Germany
\end{center}

\begin{abstract}\noindent
We calculate the next-to-leading-order QCD corrections to the perturbative
term in the operator product expansion of the spectral functions of light
tetraquark currents. By using also configuration-space methods we keep the
momentum-space four-loop calculation to a manageable level. We find that the
next-to-leading-order corrections to the perturbative term are large and can
amount to $O(100\%)$. The corrections to the corresponding Borel sum rules,
however, are small since the nonperturbative condensate contributions
dominate the Borel sum rules.
\end{abstract}

\newpage


\begin{figure}
\begin{center}
\epsfig{figure=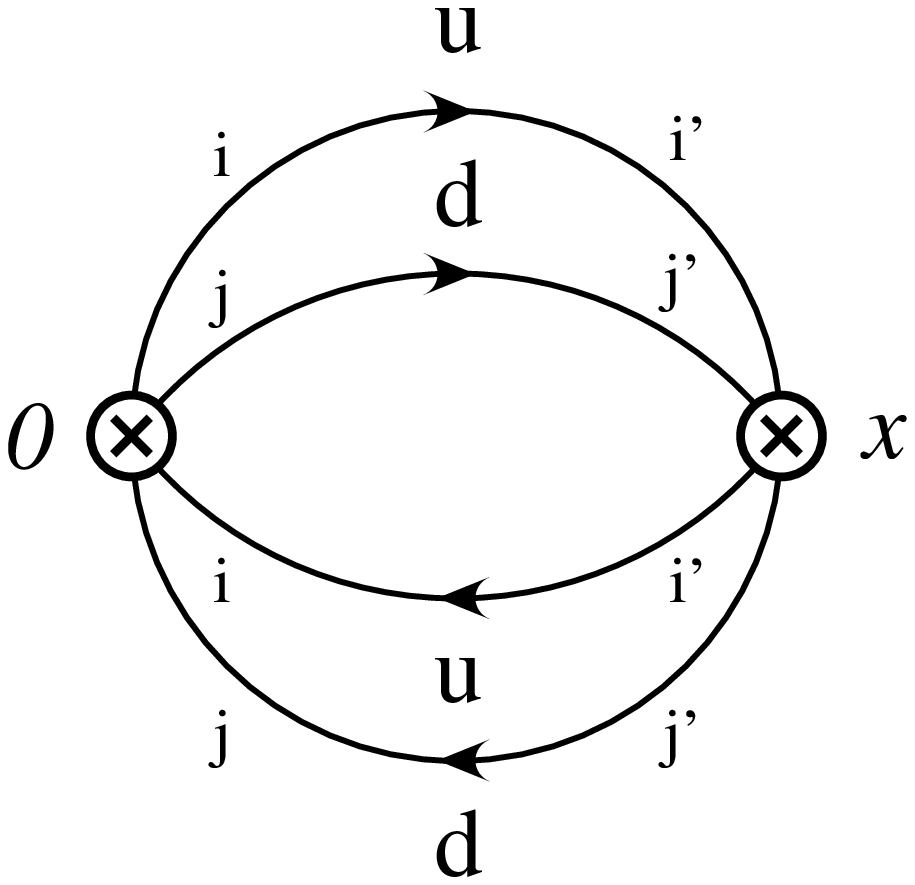, scale=0.5}
\quad
\epsfig{figure=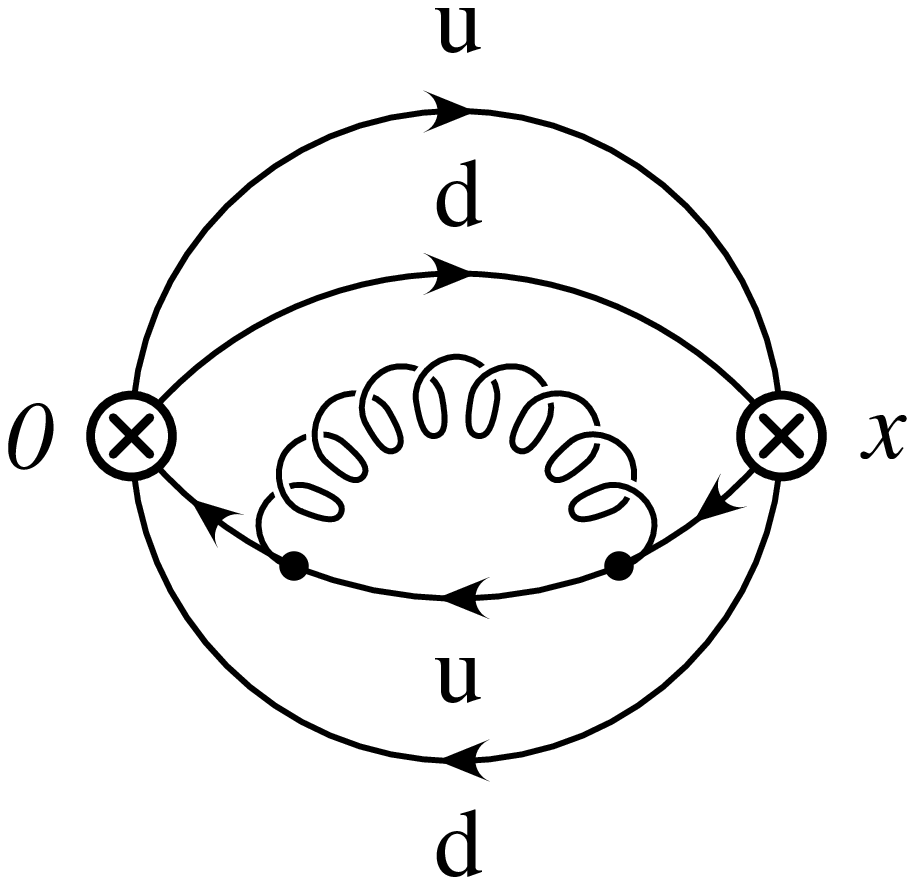, scale=0.5}
\quad
\epsfig{figure=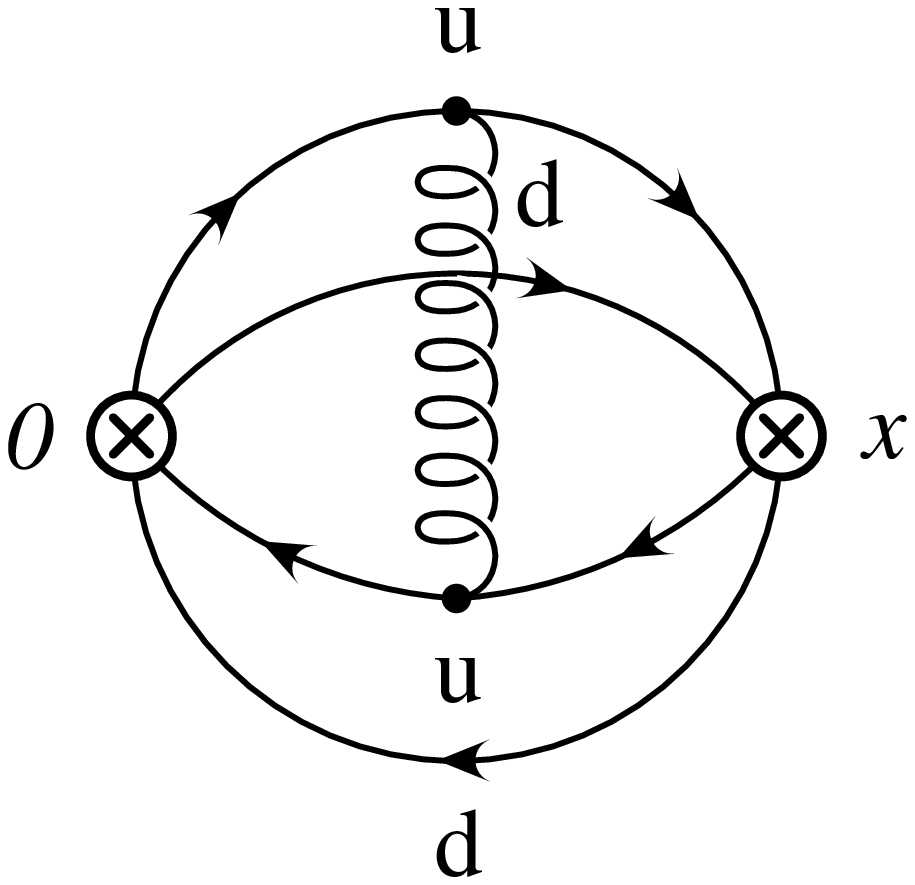, scale=0.5}\\[12pt]
(a)\kern130pt(b)\kern130pt(c)
\caption{\label{tetcor}Leading-order tetraquark correlator (a)
  and two first-order corrections (b), (c)}
\end{center}
\end{figure}

\section{Introduction}
The nonet of light scalar mesons $\sigma(600)$, $\kappa(800)$, $a_{0}(980)$
and $f_{0}(980)$ are prime candidates for the long-sought-after light 
tetraquark states. Their mass ordering $m_\sigma<m_\kappa<m_{a_0,f_0}$
precludes a simple $(\bar qq)$ interpretation~\cite{Jaffe:1976ig,Jaffe:1976ih}.
Also, in a $(\bar qq)$ picture, their masses are expected to lie above $1\GeV$
contrary to experiment~\cite{Close:2002zu}. The spectrum of these four light
scalar meson states fits perfectly into a picture where they are viewed as
$L=0$ bound states of colour-, flavour- and spin-antisymmetric light diquarks
and antidiquarks~\cite{Jaffe:1976ig,Jaffe:1976ih}. In this picture one obtains
a nonet of light scalar mesons composed of the states
$a^0_0(I=1,I_3=0)/f_0(I=0)=([su][\bar s\bar u]\mp[sd][\bar s\bar d])/\sqrt2$,
$\sigma(I=0)=[ud][\bar u\bar d]$, $\kappa^0=[ud][\bar u\bar s]$,
$\bar\kappa^0=[us][\bar u\bar d]$ and the corresponding charged states
$a^\pm_0$ and $\kappa^\pm$ in which the degeneracy of the two states $a^0_0$
and $f_0$ is natural and in which one obtains the above mass hierarchy
$m_\sigma<m_\kappa<m_{a_0,f_0}$ (see e.g.\
Refs.~\cite{Jaffe:2003sg,Maiani:2004uc}). Recently S.~Weinberg has
investigated tetraquarks in the large-$N_c$ limit of
QCD~\cite{Weinberg:2013cfa} and found the existence of light tetraquark states
to be consistent with large-$N_c$ QCD contrary to previous statements in the
literature. An interesting development was described in
Ref.~\cite{Hooft:2008we}. Instantons produce an effective six-quark vertex
which, among others, provides a mechanism for the decay $f_0(980)\to\pi\,\pi$.
 
A central theoretical issue is the need to theoretically understand the mass
pattern of the light scalar states and whether a tetraquark interpretation of
these states is able to accommodate or even predict the mass pattern of the
light scalar states. This issue has been addressed in a number of recent
theoretical investigations using the framework of QCD (Borel) sum rules to
study the properties of light tetraquark states~\cite{Brito:2004tv,%
Chen:2006hy,Chen:2007zzg,Chen:2007xr,Chen:2008iw,Wang:2005cn,Wang:2005xy,%
Wang:2006gj,Wang:2006ri,Wang:2010pn,Lee:2007mva,Pang:2009zs}. Perhaps the most
complete of these is the analysis  by Chen, Hosaka and Zhu~\cite{Chen:2007xr}.
They studied the most general form of interpolating currents including
possible mixing effects between them. In the operator product expansion they
included up to dimension-eight operators. However, in their analysis and in
previous analyses next-to-leading-order (NLO) QCD corrections to the
leading-order (LO) perturbative term were not included. As has been emphasized
by Zhang {\it et al.\/} one needs to calculate the $\alpha_s$ corrections to
the current correlators in order to make the sum rule analysis reliable and
predictable~\cite{Zhang:2007bt}. In momentum space ($p$ space) the NLO
corrections to the light tetraquark current correlators or spectral functions
require the calculation of massless four-loop diagrams which is not simple.
However, if one also uses configuration space ($x$-space) techniques the task
becomes simpler. This has been demonstrated in two previous papers where we
have calculated the five-loop NLO corrections to pentaquark current
correlators using also $x$-space techniques~\cite{Groote:2006sy,Groote:2011my}.
The main idea of the $x$-space calculation is to first calculate two $x$-space
modules corresponding to NLO propagator and dipropagator corrections and then
to insert the modules into the full correlator diagram. In this way the
calculation of radiative corrections to multiquark correlators amounts to 
purely algebraic manipulations.

The purpose of this paper is twofold. First we expound on the calculation of
the two $O(\alpha_s)$ modules that go into the modular approach to the
calculation of radiative corrections to the current--current correlators of 
multiquark currents. As a new feature compared to~\cite{Groote:2006sy,%
Groote:2011my} we show, by using a general $R_\xi$ gauge, that the sum of the
two modules is gauge invariant when sandwiched between colour-neutral states
(mesons, baryons, tetraquarks and pentaquarks). Second, we present explicit 
results on the radiative corrections of light tetraquark current correlators
for the two sets of five tetraquark currents (scalar, vector, tensor, axial
vector and pseudoscalar, each for flavour-symmetric and -antisymmetric diquark
configurations) that have been investigated in Refs.~\cite{Chen:2006hy,%
Chen:2007zzg,Chen:2007xr,Chen:2008iw}. We also present NLO results on all
possible nondiagonal correlators. 

Depending on the choice of tetraquark currents the radiative corrections to
the LO perturbative term can amount to up to $132\%$ at $q^2=1\GeV^2$ for the
tensor current to be discussed later on. As a further exemplary case we
consider the spectral density corresponding to the current correlator of a
particular linear combination of the axial and vector tetraquark current
$\eta_1^\sigma$ considered in Ref.~\cite{Chen:2007xr}. The mixed current was
found to be an optimal interpolating current with a good Borel window for the 
$\sigma$-meson tetraquark current~\cite{Chen:2007xr}. For the corresponding
spectral function we list the LO perturbative result and the NLO correction
which is calculated in this paper. One has
\begin{equation}
\rho^\sigma_1(s)=\frac{s^4}{11520\pi^6}\bigg\{1+\frac{\alpha_s}\pi
  \bigg(\frac{7+6\sqrt2}4\ln\pfrac{\mu_\msbar^2}s+\frac{1381+15\sqrt2}{180}
  \bigg)\bigg\}
\end{equation}
Using $\mu_\msbar^2=1\GeV^2$ the NLO corrections can be seen to amount to a
$55\%$ upward correction to the LO term at $q^2=1\GeV^2$. Including also the
nonperturbative contributions and using the central values for the masses and
condensates from Ref.~\cite{Chen:2007xr} one finds
\begin{eqnarray}\label{condensate}
\lefteqn{\rho^\sigma_1(s=1\GeV^2)\ =\ \Big(9.03\,({\rm LO})+4.94\,({\rm NLO})}
  \nonumber\\&&
-0.042\,[d]^2+(83.89-5.80)\,[d]^4-0.045\,[d]^6-0.486\,[d]^8\Big)
\cdot 10^{-8}\GeV^8
\end{eqnarray}
The spectral function is dominated by the dimension-four gluon condensate
contribution $\langle g^2GG\rangle$ listed as the first dimension-four term in 
Eq.~(\ref{condensate}). Such a large contribution does not appear to be very
natural.  We mention that the value of the gluon condensate has not yet been
calculated from first principles but is obtained from fits to QCD sum rules.
For example, the authors of Refs.~\cite{Bordes:2005wv,Dominguez:2006ct,%
Bodenstein:2011hm} found $\langle g^2GG\rangle=(0.47\pm0.47)\GeV^4$ compared to
$\langle g^2GG\rangle=(0.48\pm 0.14)\GeV^4$ given in Ref.~\cite{Narison:2002pw}
and used by Chen {\it et al.\/} in Ref.~\cite{Chen:2007xr}. This shows that a
small or even vanishing contribution of the sum of the dimension-four
condensates to the spectral function lies within a one-standard-deviation
window of the central value if the results of
Refs.~\cite{Bordes:2005wv,Dominguez:2006ct,Bodenstein:2011hm} are used. 

\subsection{\label{inter}Interpolating currents}
For the construction of interpolating currents we refer to the detailed
presentation in Ref.~\cite{Chen:2007xr}. Following these authors we obtain two
sets of five currents each for the flavour-symmetric and -antisymmetric
diquark--antidiquark states. For conciseness we specify our currents to the
$(ud\bar u\bar d)$ sector with hypercharge $Y=0$ which make up the
$(ud\bar u\bar d)$ components of the $\sigma,f_0,a_0$ currents. In
Ref.~\cite{Chen:2007xr} one can find a detailed discussion about the flavour
composition of the various tetraquark currents. 

For the flavour-antisymmetric case one has the five currents
\begin{eqnarray}\label{flasym}
J_{S_3}^\sigma&=&(u^{iT}C\gamma_5d^j)(\bar u_i\gamma_5C\bar d_j^T
  -\bar u_j\gamma_5C\bar d_i^T),\nonumber\\
J_{V_3}^\sigma&=&(u^{iT}C\gamma^\mu\gamma_5d^j)
  (\bar u_i\gamma_\mu\gamma_5C\bar d_j^T
  -\bar u_j\gamma_\mu\gamma_5C\bar d_i^T),\nonumber \\
J_{T_6}^\sigma&=&(u^{iT}C\sigma^{\mu\nu}d^j)(\bar u_i\sigma_{\mu\nu}C\bar d_j^T
  +\bar u_j\sigma_{\mu\nu}C\bar d_i^T),\nonumber \\
J_{A_6}^\sigma&=&(u^{iT}C\gamma^\mu d^j)(\bar u_i\gamma_\mu C\bar d_j^T
  +\bar u_j\gamma_\mu C\bar d_i^T),\nonumber \\
J_{P_3}^\sigma&=&(u^{iT}Cd^j)(\bar u_iC\bar d_j^T
  -\bar u_jC\bar d_i^T),
\end{eqnarray}
where $\sigma^{\mu\nu}=\frac i2[\gamma^\mu,\gamma^\nu]$ (see also Appendix~A).
The lower index on the currents marks the colour multiplicity of the diquark
state which is given by the antisymmetric (symmetric) colour representations
in the decomposition ${\bf 3}\otimes{\bf 3}\Rightarrow{\bf\bar 3_a}$
(${\bf 3}\otimes{\bf 3}\Rightarrow{\bf\bar 6_s}$) and
${\bf\bar 3}\otimes{\bf\bar 3}\Rightarrow{\bf 3_a}$
(${\bf\bar 3}\otimes{\bf\bar 3}\Rightarrow{\bf\bar 6_s}$). We mention that the
mixed current correlator discussed above corresponds to the mixed current
\begin{equation}\label{mixed}
J_1^\sigma=\cos\theta J_{A_6}^\sigma+\sin\theta J_{V_3}^\sigma
\end{equation}
with $\tan\theta=\sqrt2$.

For the flavour-symmetric case one has the five currents 
\begin{eqnarray}\label{flsym}
J_{S_6}^\sigma&=&(u^{iT}C\gamma_5d^j)(\bar u_i\gamma_5C\bar d_j^T
  +\bar u_j\gamma_5C\bar d_i^T),\nonumber\\
J_{V_6}^\sigma&=&(u^{iT}C\gamma^\mu\gamma_5d^j)
  (\bar u_i\gamma_\mu\gamma_5C\bar d_j^T
  +\bar u_j\gamma_\mu\gamma_5C\bar d_i^T),\nonumber \\
J_{T_3}^\sigma&=&(u^{iT}C\sigma^{\mu\nu}d^j)(\bar u_i\sigma_{\mu\nu}C\bar d_j^T
  -\bar u_j\sigma_{\mu\nu}C\bar d_i^T),\nonumber \\
J_{A_3}^\sigma&=&(u^{iT}C\gamma^\mu d^j)(\bar u_i\gamma_\mu C\bar d_j^T
  -\bar u_j\gamma_\mu C\bar d_i^T),\nonumber \\
J_{P_6}^\sigma&=&(u^{iT}Cd^j)(\bar u_iC\bar d_j^T
  +\bar u_jC\bar d_i^T).
\end{eqnarray}
Except for the LO term the $O(\alpha_s)$ perturbative contributions to the
flavour-symmetric and -antisymmetric correlators are not always simply
related. The currents in Eqs.~(\ref{flasym}) and~(\ref{flsym}) are built
from diquark--antidiquark components. One can also construct the tetraquark
currents from meson--meson components. However, the meson--meson currents do
not lead to new tetraquark configurations since the two representations are
related by a Fierz transformation (see Ref.~\cite{Chen:2007xr} and Appendix~B).

For an understanding and illustration of the modular approach it is
sufficient to discuss a simplified form of the scalar current given by
\begin{eqnarray}\label{eq:skalaarvool}
J_S(x)&=&\left(u^{iT}(x)C\gamma_5d^j(x)\right)
\left(\bar u_i(x)\gamma_5C\bar d_j^T(x)\right)\nonumber\\
  &=&\delta_i^k\delta_j^l\left(u^{iT}(x)C\gamma_5d^j(x)\right)
  \left(\bar u_k(x)\gamma_5C\bar d_l^T(x)\right),
\end{eqnarray}
involving only the first part of the scalar current in Eq.~(\ref{flasym}). 
It is not difficult to reinstate the colour-symmetrized/-antisymmetrized form
of the current at the final stages of the calculations.

\subsection{The correlator}
The correlation function\footnote{In the following we use the synonym
``correlator''.} is defined as the vacuum expectation value of the 
time-ordered product of two currents, i.e.\
\begin{equation}\label{top}
\Pi(x)=\langle 0|\mathcal{T}\{J(x)\bar J(0)\}|0\rangle.
\end{equation}
If the current describes a boson (meson or tetraquark), one has 
$\bar J(0)=J^\dagger(0)$ while in case of a fermion (baryon or pentaquark),
one has $\bar J(0)=J^\dagger(0)\gamma^0$. 

The correlator in Eq.~(\ref{top}) is defined in $x$ space. It can be 
transformed to $p$ space by a Fourier transformation with the result  
\begin{equation}
\Pi(q)=i\int\Pi(x)e^{iqx}d^4x
  =i\int\langle 0|\mathcal{T}\{J(x)\bar{J}(0)\}|0\rangle e^{iqx}d^4x,
\end{equation}
where, for the moment, we work in $D=4$ dimensions. The optical theorem
relates the $p$-space correlator to the spectral density
\begin{equation}
\rho(s)=\frac1{2\pi i}\Disc\Pi(q)|_{q^2=-s},
\end{equation}
where the discontinuity $\Disc f(p)$ is defined by 
(see e.g.\ Ref.~\cite{Groote:2005ay})
\begin{equation}
\Disc f(p):=f(pe^{i0})-f(pe^{-i0}),\qquad 
e^{\pm i0}=\lim_{\eps\to 0}e^{\pm i\eps},\ \eps>0.
\end{equation}
Vice versa, for a given spectral density, the correlator can be
reconstructed by using
\begin{equation}\label{eq:korrelaator_v1}
\Pi(q)=\int_0^\infty\frac{\rho(s)ds}{s+q^2}.
\end{equation}
For the simplified scalar current in Eq.~(\ref{eq:skalaarvool}) the tetraquark
correlator in $x$ space reads
\begin{equation}\label{xspace}
\Pi(x)=\Tr\Big(S_u(x)^i_{i'}\gamma_5S_d(x)^j_{j'}\gamma_5\Big)
  \Tr\Big(S_u(-x)^{i'}_i\gamma_5S_d(-x)^{j'}_j\gamma_5\Big)=\frac9{\pi^8x^{12}}
\end{equation}
where we have made use of the $x$-space propagator given by
$S(x)^i_{i'}=\delta^i_{i'}S_0(-x^2)x^\mu\gamma_\mu$ with
$S_0(x^2)=(2\pi^2x^4)^{-1}$ for $D=4$ (cf.\ Eq.~(\ref{S10x})). The first trace
in Eq.~(\ref{xspace}) contains two quark propagators with a positive $x$-space
argument while the second trace contains two antiquark propagators
corresponding to quark propagators with a negative $x$-space argument. The
general rule is that an antiquark propagator carries an extra minus sign. Note
that the colour trace in Eq.~(\ref{xspace}) connects quarks/antiquarks in the
two different Dirac traces. 

\section{Propagator and dipropagator modules}
The result in Eq.~(\ref{xspace}) reflects a very general property of massless 
correlators represented by sunrise-type diagrams: in $x$ space they are
obtained by a product of single $x$-space propagators. The corresponding 
$p$-space calculation is far more difficult since one would have to perform a 
genuine three-loop calculation. This observation sets the strategy for the
evaluation of the radiative corrections to the tetraquark correlator: do most
of the calculation in $x$ space. In detail, we first calculate the radiative
corrections to a single propagator and the dipropagator in $p$ space (see
Fig.~\ref{Flo:parandid1}). We shall refer to these two corrections as the
propagator and dipropagator modules. In the two modules the Dirac and colour
indices are left open. We then Fourier transform the two modules to $x$ space.
Next we assemble the $x$-space tetraquark correlator from these two modules
augmented by free propagators as shown in in Fig.~\ref{Flo:parandid1}. The
assembly is simple in $x$ space since the free propagators are linked to the
modules in product form. One then does the appropriate Dirac and colour
contractions according to the specific current being investigated. Finally, we
Fourier transform the $x$-space tetraquark correlator back to $p$ space.

In the following section we first calculate the $p$-space propagator and the
dipropagator corrections using traditional momentum integration methods. The
propagator and the dipropagator corrections are then transformed to $x$ space.

\subsection{The propagator correction}
\begin{figure}
\begin{center}
\epsfig{figure=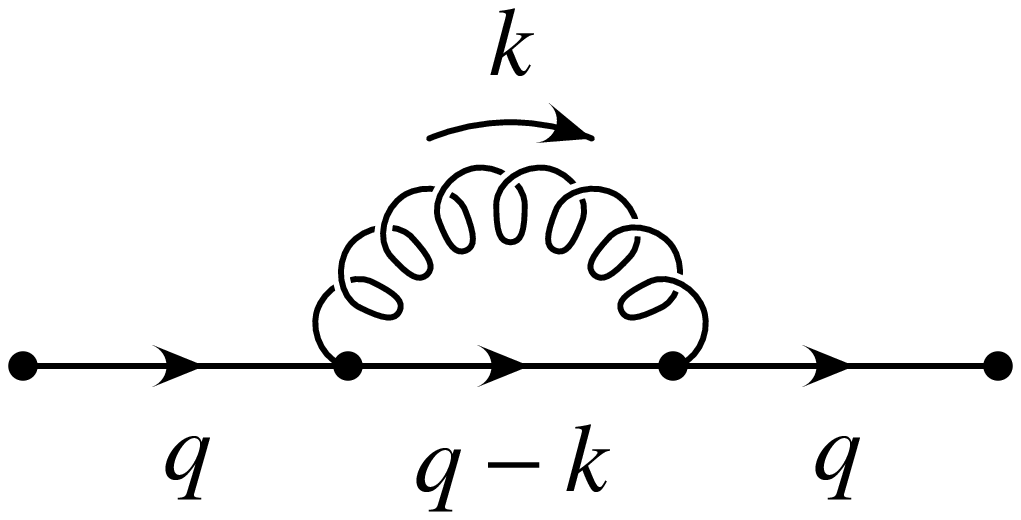, scale=0.5}\qquad
\epsfig{figure=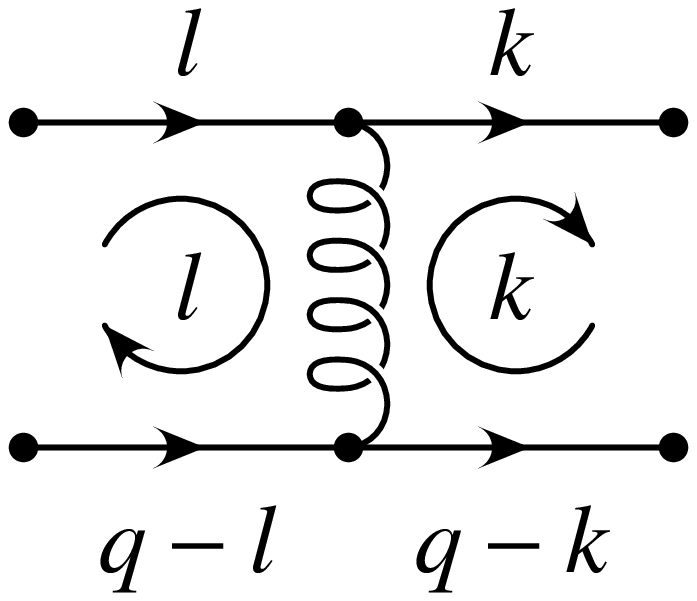, scale=0.5}\\[12pt]
\strut\qquad(a)\kern130pt(b)
\caption{Propagator correction (a) with loop momentum $k$ and dipropagator
correction (b) with loop momenta $l$ and $k$}
\label{Flo:parandid1}
\end{center}
\end{figure}

For illustrative reasons we begin by considering the LO massless propagator
in $p$ space which takes the familiar form
\begin{equation}\label{pspace}
S_1^0(q)=\frac{i}{\slq}=-iq_\mu\gamma^\mu(-q^2)^{-1}.
\end{equation}
In order to obtain the corresponding LO $x$-space propagator we have to take
the Fourier transform of the propagator in Eq.~(\ref{pspace}). Since we are
working in dimensional regularization one needs to make use of the
$D$-dimensional Fourier transform ($D=4-2\eps$). The relevant $D$-dimensional
transformation formulas are collected in Appendix~C. One obtains 
\begin{equation}\label{S10x}
S_1^0(x)=S_0(-x^2)x_\mu\gamma^\mu,
\end{equation}
where we have factored out a frequently occurring function $S_0(-x^2)$
defined by
\begin{equation}
S_0(-x^2)=\frac{\Gamma(2-\eps)}{2(4\pi)^{2-\eps}}
  \left(-\frac{x^2}4\right)^{\eps-2}
  =\frac{\Gamma(2-\eps)}{2\pi^{2-\eps}(-x^2)^{2-\eps}}.
\end{equation}
In Feynman gauge and in $p$ space the propagator correction (see
Fig.~\ref{Flo:parandid1}(a)) reads\footnote{In dimensional regularization the
strong charge $\tilde g_s$ has a mass dimension which will be absorbed into the
renormalization scale such that one remains with a dimensionless renormalized
charge $g_s=g_s(\mu)$.}
\begin{equation}\label{propcor}
S_1^1(q)=\frac{i}\slq\int\dDk(-i\tilde g_s\gamma^\alpha T_a)
  \frac{i}\slk(-i\tilde g_s\gamma^\beta T_b)\frac{i}\slq
  \,\frac{-ig_{\alpha\beta}\delta_{ab}}{(q-k)^2}.
\end{equation}
In Eqs.~(\ref{pspace})--(\ref{propcor}) we have suppressed the colour index 
dependence $\delta_i^j$.

Let us briefly comment on the gauge dependence of our results. In a general
$R_\xi$ gauge the gluon propagator reads
\begin{equation}
D_{\alpha\beta}(k)=\frac{i}{k^2}\left(-g_{\alpha\beta}
  +(1-\xi)\frac{k_{\alpha}k_{\beta}}{k^2}\right).
\end{equation}
The Feynman gauge used in Eq.~(\ref{propcor}) corresponds to the choice 
$\xi=1$. As shown in Appendix~D, the gauge dependence drops out in the sum of
the propagator and dipropagator corrections when sandwiched between
colour-neutral states.

Returning to Eq.~(\ref{propcor}) we proceed with the Feynman gauge calculation
and obtain
\begin{eqnarray}
\label{propcorp}
S_1^1(q)&=&(D-2)\tilde g_s^2T_aT_a\frac i\slq\int\dDk
  \frac\slk{k^2(q-k)^2}\frac i\slq.
\end{eqnarray}
It is convenient to define the dimensionless one-loop two-point integrals 
$G(n_1,n_2)$ through
\begin{equation}
\frac{i}{(4\pi)^{D/2}}(-q^2)^{D/2-n_1-n_2}G(n_1,n_2)
  :=\int\dDk\frac1{(-k^2)^{n_1}(-(q-k)^2)^{n_2}}.
\end{equation}
After setting the tadpole contributions to zero, the integral in
Eq.~(\ref{propcorp}) can be expressed in terms of the standard integral
$G(1,1)$ given by 
\begin{equation}
\label{G11}
G(1,1)=\frac{\Gamma(\eps)\Gamma^2(1-\eps)}{\Gamma(2-2\eps)}
  =\frac{\Gamma(1+\eps)\Gamma^2(1-\eps)}{\eps\Gamma(2-2\eps)}=:\frac G\eps.
\end{equation}
The one-loop integral $G(1,1)$ is divergent. In Eq.~(\ref{G11}) we have
introduced the factor $G=\Gamma(1+\eps)\Gamma^2(1-\eps)/\eps\Gamma(2-2\eps)$
because we want to absorb the $\Gamma$ factors in Eq.~(\ref{G11}) into the
definition of the renormalization scale. We shall refer to this scheme as 
the $G$ scheme. The 
relation to the (modified) minimal substraction scheme is given by 
\begin{equation}
\mu_G^{2\eps}=G(4\pi\mu_\ms^2)^\eps,\qquad
\mu_G^2\approx e^2\mu_\msbar^2=4\pi e^{2-\gamma_E}\mu_\ms^2.
\end{equation}
The corrected NLO $p$-space propagator finally reads 
\begin{equation}
S_1^1(q)=-\frac{\alpha_sC_F}{4\pi}\left(\frac1\eps-1\right)
 \left(-\frac{q^2}{\mu_G^2}\right)^{-\eps}\frac{i}\slq.
\end{equation}
With the help of Eq.~(\ref{eq:konfruum2}) in Appendix~C (with $\alpha=1+\eps$)
one obtains the corresponding $x$-space result
\begin{equation}
S_1^1(x)=-S_0(-x^2)\frac{\alpha_sC_F}{4\pi}\left(\frac1\eps+O(\eps)\right)
  (-\mu_x^2x^2)^\eps x_\mu\gamma^\mu.
\end{equation}
We have introduced an $x$-space scale $\mu_x$ as a new renormalization 
scale,
defined by
\begin{equation}
\mu_x^{2\eps}=\frac{\Gamma(1-\eps)}{(4\pi)^{-\eps}}
  \pfrac14^\eps\mu_\ms^{2\eps},\qquad
  4\mu_x^2=4\pi e^{\gamma_E}\mu_\ms^2+O(\eps)
  =e^{2\gamma_E}\mu_\msbar^2+O(\eps).
\end{equation}
Altogether the $O(\alpha_s)$ $x$-space propagator reads
\begin{equation}\label{eq:prop_parand}
S_1(x)=S_0(-x^2)\left\{1-\frac{\alpha_sC_F}{4\pi}
  \left(\frac1\eps+O(\eps)\right)(-\mu_x^2x^2)^\eps+O(\alpha_s^2)\right\}
  x_\mu\gamma^\mu.
\end{equation}
As expected, the propagator correction has the spatial structure of the LO 
term in Eq.~(\ref{S10x}).

\subsection{The dipropagator correction}
In order to familiarize the reader with the calculational procedure and the 
notation, we start our discussion with the calculation of the uncorrected
dipropagator. In $p$ space the uncorrected dipropagator consists of a single
loop integral where the two pairs of Dirac and colour indices are left open
and uncontracted. The one-loop integral reads (colour indices are suppressed) 
\begin{equation}
S_2^0(q)=\int\dDk\left(\frac{i}\slk\otimes\frac{i}{\slq-\slk}\right)
  =\int\dDk\frac{k_{\mu}(q-k)_{\nu}}{k^2(q-k)^2}(\gamma^\mu\otimes\gamma^\nu)
  =I_{\mu\nu}^0(q)(\gamma^\mu\otimes\gamma^\nu).
\end{equation}
Expanding the tensor integral $I_{\mu\nu}^0$ into the two covariants
$q^2g_{\mu\nu}$ and $q_\mu q_\nu$, one has
\begin{equation}\label{diprop}
I_{\mu\nu}^0(q)=\int\dDk\frac{k_\mu(q-k)_\nu}{k^2(q-k)^2}
  =A^0q_\mu q_\nu+B^0q^2g_{\mu\nu}.
\end{equation}
By contracting Eq.~(\ref{diprop}) with $g^{\mu\nu}$ and $q^{\mu}q^{\nu}$,
calculating the resulting scalar integrals, dropping tadpole contributions 
and solving for $A^0$ and $B^0$ one obtains 
\begin{equation}
A^0=\frac{-iG(1,1)(D-2)}{4(4\pi)^{D/2}(D-1)}(-q^2)^{D/2-2},\qquad
B^0=\frac{-iG(1,1)}{4(4\pi)^{D/2}(D-1)}(-q^2)^{D/2-2},
\end{equation}
where the scalar integral $G(1,1)$ is listed in Eq.~(\ref{G11}). Altogether
one has
\begin{equation}\label{eq:diprop2}
S_2^0(q)=\frac{-iG(1,1)}{4(4\pi)^{D/2}(D-1)}(-q^2)^{D/2-2}
  \Big((D-2)q_{\mu}q_{\nu}+q^2g_{\mu\nu}\Big)(\gamma^\mu\otimes\gamma^\nu).
\end{equation}
We then Fourier transform $S_2^0(q)$ to $x$ space using the results in 
Appendix~C (Eq.~(\ref{eq:konfruum4}) with $\alpha=2-D/2=\eps$, 
$A=D-2=2(1-\eps)$ and $B=1$). One obtains
\begin{eqnarray}\label{diuncor}
S_2^0(x)&=&\frac{G(1,1)\Gamma(D-1)}{32(4\pi)^D(D-1)\Gamma(2-D/2)}
  \left(-\frac{x^2}4\right)^{-D}\Big(-2(D-1)(D-2)x_\mu x_\nu\Big)
  (\gamma^\mu\otimes\gamma^\nu)\nonumber\\
  &=&(S_0(-x^2))^2x_\mu x_\nu(\gamma^\mu\otimes\gamma^\nu)
  \ =\ S_1^0(x)\otimes S_1^0(x).
\end{eqnarray}
where $S_1^0(x)$ is defined in Eq.~(\ref{S10x}). The factorized result in the
second line of Eq.~(\ref{diuncor}) is a special case of the general $x$-space
result for a massless $n$-loop sunrise diagram written down in
Refs.~\cite{Groote:2005ay,Groote:1998ic,Groote:1998wy,Groote:1999cx}.

In order to calculate the NLO dipropagator correction we start again in 
$p$ space. A symbolic representation of the corresponding two-loop Feynman 
diagram is shown in Fig.~\ref{Flo:parandid1}(b). The endpoints of the momentum
lines in the initial and final states have not been joined together in order
to symbolize the fact that the colour and Dirac indices in the diagram are
left open. The two-loop correction to the dipropagator is given by the
twofold integral
\begin{eqnarray}\label{dipropmom}
S_2^1(q)&=&\int\dDk\dDl\left(\frac{i}\slk(-i\tilde g_s\gamma^\alpha T_a)
  \frac{i}\sll\otimes\frac{i}{\slq-\slk}(-i\tilde g_s\gamma^\beta T_b)
  \frac{i}{\slq-\sll}\right)\frac{-ig_{\alpha\beta}\delta_{ab}}{(k-l)^2}
  \nonumber\\
  &=&(T_a\otimes T_a)\tilde g_s^2(\gamma^\mu\gamma^\alpha\gamma^\nu\otimes
  \gamma_{\mu'}\gamma_\alpha\gamma_{\nu'})I^{\mu'\nu'}_{\mu\nu}(q),
\end{eqnarray}
where
\begin{equation}
I^{\mu'\nu'}_{\mu\nu}(q)
  =i\int\dDk\dDl\frac{k_\mu l_\nu(q-k)^{\mu'}(q-l)^{\nu'}}{k^2
  l^2(q-k)^2(q-l)^2(k-l)^2}.
\end{equation}
The integral $I^{\mu'\nu'}_{\mu\nu}(q)$ can be seen to be symmetric under the
{\em simultaneous\/} interchange of $\mu\leftrightarrow\nu$ and 
$\mu'\leftrightarrow\nu'$. It is therefore expedient to split the gamma matrix
string $\gamma^\mu\gamma^\alpha\gamma^\nu$ (and, accordingly,
$\gamma_{\mu'}\gamma_\alpha\gamma_{\nu'}$) into its $\mu\leftrightarrow\nu$
and $\mu'\leftrightarrow\nu'$ symmetric and antisymmetric parts,
\begin{equation}
\gamma^\mu\gamma^\alpha\gamma^\nu
  =\frac12(\gamma^\mu\gamma^\alpha\gamma^\nu
  +\gamma^\nu\gamma^\alpha\gamma^\mu)
  +\frac12(\gamma^\mu\gamma^\alpha\gamma^\nu
  -\gamma^\nu\gamma^\alpha\gamma^\mu)
  =\gamma^{(\mu}\gamma^\alpha\gamma^{\nu)}
  +\gamma^{[\mu}\gamma^\alpha\gamma^{\nu]}.
\end{equation}
One then remains with
\begin{eqnarray}\label{symasym}
\lefteqn{(\gamma^\mu\gamma^\alpha\gamma^\nu\otimes
  \gamma_{\mu'}\gamma_\alpha\gamma_{\nu'})I^{\mu'\nu'}_{\mu\nu}(q)\ =}
  \nonumber\\[7pt]
  &=&(\gamma^{(\mu}\gamma^\alpha\gamma^{\nu)}\otimes
  \gamma_{(\mu'}\gamma_\alpha\gamma_{\nu')})I^{(\mu'\nu')}_{(\mu\nu)}(q)
  +(\gamma^{[\mu}\gamma^\alpha\gamma^{\nu]}\otimes
  \gamma_{[\mu'}\gamma_\alpha\gamma_{\nu']})I^{[\mu'\nu']}_{[\mu\nu]}(q).\qquad
\end{eqnarray}
The symmetric--symmetric contribution in Eq.~(\ref{symasym}) will be dealt
with by making use of the $D$-dimensional identity
$\gamma^{(\mu}\gamma^\alpha\gamma^{\nu)}=(g^{\mu\alpha}g^{\nu\beta}
+g^{\mu\beta}g^{\nu\alpha}-g^{\mu\nu}g^{\alpha\beta})\gamma_\beta$ and the
corresponding identity for $\gamma_{(\mu'}\gamma_{\alpha}\gamma_{\nu')}$.

The contraction $(\gamma^{(\mu}\gamma^\alpha\gamma^{\nu)}\otimes
  \gamma_{(\mu'}\gamma_\alpha\gamma_{\nu')})\,I^{(\mu'\nu')}_{(\mu\nu)}(q)$ 
leads to a number of second-rank two-loop tensor integrals which can be 
reduced to scalar integrals using standard techniques. In order to minimize
calculational mistakes, all necessary manipulations have been done by computer
algebra programs.\footnote{The reduction to scalar integrals is performed by
integration-by-parts techniques~\cite{Chetyrkin:1981qh}. The
integration-by-parts method under the name {\tt recursor} is originally 
written in {\tt Reduce} and is translated by us for use under {\tt MATHEMATICA}
(for an overview see e.g.\ Ref.~\cite{Grozin:2011mt}).} The required set of
scalar two-loop integrals needed in this application are defined
by~\cite{Broadhurst:1991fi}
\begin{eqnarray}\label{scalar}
\lefteqn{\frac{-1}{(4\pi)^{D}}(-q^2)^{D-n_1-n_2-n_3-n_4-n_5}
  G(n_1,n_2,n_3,n_4,n_5)\ :=}\nonumber\\
  &=&\int\dDk\dDl\frac1{(-k^2)^{n_1}(-l^2)^{n_2}(-(q-k)^2)^{n_3}
  (-(q-l)^2)^{n_4}(-(k-l)^2)^{n_5}}.\qquad
\end{eqnarray}
Their solution has been given in Ref.~\cite{Broadhurst:1991fi}.

The symmetric--symmetric contribution to the dipropagator correction can be
represented in the form
\begin{equation}
\label{symsym}
S_2^{S1}(q)=(T_a\otimes T_a)(I_S)_\beta{}^{\beta'}(q)
  (\gamma^\beta\otimes\gamma_{\beta'}),
\end{equation}
where
\begin{equation}
(I_S)_\beta{}^{\beta'}(q)=A_1(q^2)q_\beta q^{\beta'}
  +B_1(q^2)q^2g_\beta^{\beta'},
\end{equation}
and where
\begin{equation}
A_1(q^2)=\frac{i\tilde g_s^2G^2(-q^2)^{-2\eps}}{(4\pi)^{4-2\eps}\eps^2}
  \pfrac{1-5\eps}{12},\qquad
B_1(q^2)=\frac{i\tilde g_s^2G^2(-q^2)^{-2\eps}}{(4\pi)^{4-2\eps}\eps^2}
  \pfrac{7+13\eps}{24}.
\end{equation}

Next we turn to the antisymmetric--antisymmetric contribution in
Eq.~(\ref{symasym}) whose structure can be further specified by noting that
the integral $I_{[\mu\nu]}^{[\mu'\nu']}(q)$ is {\em separately\/} antisymmetric
under the exchange $\mu\leftrightarrow\nu$ and $\mu'\leftrightarrow\nu'$. The
integral $I_{[\mu\nu]}^{[\mu'\nu']}(q)$ can thus be expanded into two
corresponding tensors built from the metric tensor and the outer momentum $q$
that possess this antisymmetry. One therefore has\footnote{We mention in
passing that in $D=4$ dimensions the antisymmetric Dirac string can be
simplified by using the $D=4$ identity
$\gamma^{[\mu}\gamma^{\alpha}\gamma^{\nu]}=
\frac i2\gamma_5\epsilon^{\mu\alpha\nu\beta}\gamma_\beta$. Using the same
identity for $\gamma_{[\mu'}\gamma_{\alpha}\gamma_{\nu']}$ one would end up
with the contraction
$\epsilon^{\mu\alpha\nu\beta}\epsilon_{\mu'\alpha\nu'\beta'}$ leading to a sum
of products of three metric tensors. In this case one would again have
second-order tensor integrals as in the symmetric--symmetric case but now
multiplying the Dirac structure
$\gamma_5\gamma^\beta\otimes\gamma_5\gamma_{\beta'}$. However, since we are
working within dimensional regularization, we cannot make use of the above
identity.}
\begin{equation}
I_{[\mu\nu]}^{[\mu'\nu']}(q)
  =\Big(q_\mu(g_\nu^{\nu'}q^{\mu'}-g_\nu^{\mu'}q^{\nu'})
  +q_\nu(g_\mu^{\mu'}q^{\nu'}-g_{\mu}^{\nu^{'}}q^{\mu'})\Big)R_1
  +(g_\mu^{\mu'}g_\nu^{\nu'}-g_\mu^{\nu'}g_\nu^{\mu'})R_2.
\end{equation}
Using again standard techniques one obtains the contributions of the
antisymmetric--antisymmetric part in terms of a set of fourth-order tensor
integrals which can again be reduced to the two-loop scalar integrals in
Eq.~(\ref{scalar}).

Similar to Eq.~(\ref{symsym}) the antisymmetric--antisymmetric contribution to
the dipropagator correction can be written in the form
\begin{equation}\label{asymasym}
S_2^{A1}(q)=(T_a\otimes T_a)(I_A)_\beta{}^{\beta'}(q)
  (\gamma^{[\mu}\gamma^\beta\gamma^{\nu]}
  \otimes\gamma_{[\mu'}\gamma_{\beta'}\gamma_{\nu']})
\end{equation}
with
\begin{equation}
(I_A)_\beta{}^{\beta'}(q)=A_3(q^2)q_\beta q^{\beta'}
  +B_3(q^2)q^2g_\beta^{\beta'},
\end{equation}
and 
\begin{equation}
A_3(q^2)=\frac{i\tilde g_s^2G^2(-q^2)^{-2\eps}}{(4\pi)^{4-2\eps}\eps^2}
  \pfrac1{24},\qquad
B_3(q^2)=\frac{i\tilde g_s^2G^2(-q^2)^{-2\eps}}{(4\pi)^{4-2\eps}\eps^2}
  \pfrac{1+2\eps}{48}.
\end{equation}
Adding up the symmetric--symmetric contribution in Eq.~(\ref{symsym}) and the
antisymmetric--antisymmetric contribution in Eq.~(\ref{asymasym}) the final
result in $p$ space reads 
\begin{eqnarray}\label{eq:diprop3}
S_2^1(q)&=&(T_a\otimes T_a)
  \frac{i\tilde g_s^2G^2(-q^2)^{-2\eps}}{(4\pi)^{4-2\eps}\eps^2}
  \Big\{(a_1q_\beta q^{\beta'}+b_1q^2g_\beta^{\beta'})
  (\gamma^\beta\otimes\gamma_{\beta'})\nonumber\\&&
  +(a_3q_\beta q^{\beta'}+b_3q^2g_\beta^{\beta'})
  (\gamma^{[\mu}\gamma^\alpha\gamma^{\beta]}\otimes
  \gamma_{[\mu}\gamma_\alpha\gamma_{\beta']})\Big\},
\end{eqnarray}
where
\begin{equation}
a_1=\frac1{12}(1-5\eps),\qquad 
b_1=\frac1{24}(7+13\eps),\qquad 
a_3=\frac1{24},\qquad
b_{3x}=\frac1{48}(1+2\eps).
\end{equation}
The result is then Fourier transformed to $x$ space using again the results
of Appendix~C. Together with the LO result one finally has
$S_2(x)=S_2^0(x)+S_2^1(x)$,
\begin{eqnarray}\label{eq:diprop_parand}
S_2(x)&=&(S_0(-x^2))^2\bigg\{\bigg[\bigg(1-(T_a\otimes T_a)
  \frac{\alpha_s}{4\pi}(-\mu_x^2x^2)^\eps
  \left(\frac1\eps+\frac{11}2\right)x_{\mu}x_{\nu}\nonumber\\&&
  -(T_a\otimes T_a)\frac{\alpha_s}{4\pi}(-\mu_x^2x^2)^\eps
  \left(\frac1\eps+\frac12\right)x^2g_{\mu\nu}\bigg]
  (\gamma^\mu\otimes\gamma^\nu)\nonumber\\&&
  -(T_a\otimes T_a)\frac{\alpha_s}{4\pi}(-\mu_x^2x^2)^\eps
  \left(\frac1{2\eps}+\frac14\right)x_\mu x_\nu
  (\gamma^{[\alpha}\gamma^\beta\gamma^{\mu]}\otimes
  \gamma_{[\alpha}\gamma_{\beta}\gamma^{\nu]})\bigg\}.\qquad
\end{eqnarray}
It is important to realize that the dipropagator module contains terms
which do not have the spatial structure of the LO term in
Eq.~(\ref{eq:diprop3}).

\section{Renormalization}
In as much as the divergence of the propagator can be removed by renormalizing
the wave function, one can remove the divergences of the correlator $\Pi$ by
renormalizing the currents $J$. There is an important difference, though, in
as much as the corrected correlator may contain higher-order spinor field
products that differ from the LO currents. Therefore, we must take into
account both multiplicative renormalization and additive counterterms. If
$J^{(0)}(x)$ is the LO current and $\Delta J^{(1)}(x)$ is the first-order
correction, the first-order correction of the correlator in Eq.~(\ref{top})
is given by
\begin{equation}
\Delta\Pi^{(1)}(x)=\langle 0|\mathcal{T}\{J^{(0)}(x)\Delta\bar J^{(1)}(0)
  +\Delta J^{(1)}(x)\bar J^{(0)}(0)\}|0\rangle.
\end{equation}

\subsection{Renormalization factor}
The wave-function renormalization is calculated by using the propagator
correction. The wave-function renormalization is always multiplicative since
the propagator correction does not change the structure of the current. The
renormalization factor is a perturbation series in the corresponding
subtraction scheme which consists of pure poles to every order. The
renormalization factor for the propagator correction~(\ref{eq:prop_parand})
reads
\begin{equation}
Z_2=1+\sum_{n=1}^\infty\pfrac{\alpha_s}{4\pi}^n
  \sum_{m=1}^n\frac1{\eps^m}Z_{n}^m
  =1+\frac{\alpha_s}{4\pi\eps}Z_1^1+O(\alpha_s^2)
\end{equation}
with the condition that $Z_2^{-1}S_1(x)=:S_1^r(x)$ is finite. As a result
we obtain the renormalization factor
\begin{equation}
Z_2=1-\frac{\alpha_sC_F}{4\pi\eps}=1-\frac{\alpha_s}{3\pi\eps}.
\end{equation}
Therefore, we renormalize the singularity of the propagator corrections by
multiplying the correlator with $Z_2^{-n}$,  where $n$ is the number of lines
($n=4$ for the tetraquark).

\subsection{Counterterms}
The first-order QCD dipropagator correction changes the quark spinor fields 
$q(x)$. In $x$ space the change due to gluon exchange is given by
\begin{equation}
\Delta q(x)=-ig_s\int d^Dx'S(x-x')\gamma^\mu T_aq(x')A_\mu^a(x').
\end{equation}
where $A_\mu^a(x')$ is the gluon field. In the case of the tetraquark current
one has four different species of spinor fields: $q(x)$, $\bar q(x)$,
$q^T(x)C$, and $C\bar q^T(x)$. The corresponding changes under gluon exchange
for the current at $x=0$ (the current at $x\ne 0$ is dealt with accordingly)
are given by
\begin{eqnarray}
\Delta q_i&=&-ig_s(T_a)_{ii'}\int d^4x\,S_\alpha(-x)A_\beta^a(x)
  \gamma^\alpha\gamma^\beta q_{i'}(x),\nonumber\\[7pt]
\Delta\bar q_i&=&-ig_s\int d^4x\,\bar q_{i'}(x)\gamma^\beta\gamma^\alpha
  S_\alpha(x)A_\beta^a(x)(T_a)_{i'i},\nonumber\\[7pt]
\Delta q_i^TC&=&-ig_s(T_a)_{ii'}\int d^4x\,q_{i'}^T(x)C\gamma^\beta
  \gamma^\alpha S_\alpha(x)A_\beta^a(x),\nonumber\\[7pt]
C\Delta\bar q_i^T&=&-ig_s\int d^4x\,A_\beta^a(x)S_\alpha(-x)\gamma^\alpha
  \gamma^\beta C\bar q_{i'}^T(x)(T_a)_{i'i}.
\end{eqnarray}
In the dipropagator correction the gluon fields of two spinor corrections
have to be linked to build a gluon propagator. Therefore, one has to calculate
\begin{eqnarray}
\lefteqn{-g_s^2\int d^Dx'd^Dx''S_\alpha(s_1x')D_{\mu\nu}(x'-x'')
  S_\beta(s_2x'')\ =}\nonumber\\[7pt]
  &=&is_1s_2g_s^2\int\dDk\frac{k_\alpha k_\beta}{(k^2-m_G^2)^3}g_{\mu\nu}
  \  =\ \frac{-s_1s_2\alpha_s}{16\pi\eps}g_{\alpha\beta}g_{\mu\nu}
  +\mbox{finite terms},
\end{eqnarray}
where $s_1$ and $s_2$ are the signs of the two propagator arguments. Note that
we can perform the integration over $x'$ and $x''$ because we are looking for
the UV-singular part. In $x$ space, the UV regime means locality, i.e.\ one
can replace $\psi(x')$ and $\psi(x'')$ under the integrations by $\psi(x)$. In
addition, we have used a (small) gluon mass $m_G$ in order to regularize
IR singularities. We have kept only the UV-singular terms because in order to
calculate the counterterm we stay in the same renormalization scheme, i.e.\
we do not subtract finite terms. Taking these considerations into account, the
changes for the different formal products of two spinors read
\begin{eqnarray}
\Delta(q^i\otimes q^j)
  &=&\frac{-\alpha_s}{16\pi\eps}(T_a)^i_{i'}(T_a)^j_{j'}
  (\gamma^\alpha\gamma^\mu q^{i'}\otimes
  \gamma_\alpha\gamma_\mu q^{j'}),\nonumber\\[7pt]
\Delta(q^i\otimes\bar q_j)
  &=&\frac{+\alpha_s}{16\pi\eps}(T_a)^i_{i'}(T_a)_j^{j'}
  (\gamma^\alpha\gamma^\mu q^{i'}\otimes
  \bar q_{j'}\gamma_\mu\gamma_\alpha),\nonumber\\[7pt]
\Delta(q^{iT}C\otimes q^j)
  &=&\frac{-\alpha_s}{16\pi\eps}(T_a)^i_{i'}(T_a)^j_{j'}
  (q^{i'T}C\gamma^\mu\gamma^\alpha\otimes
  \gamma_\alpha\gamma_\mu q^{j'}),\nonumber\\[7pt]
\Delta(q_{iT}C\otimes\bar q_j)
  &=&\frac{+\alpha_s}{16\pi\eps}(T_a)^i_{i'}(T_a)_j^{j'}
  (q^{i'T}C\gamma^\mu\gamma^\alpha\otimes
  \bar q_{j'}\gamma_\mu\gamma_\alpha),\nonumber\\[7pt]
\Delta(\bar q_i\otimes q^j)
  &=&\frac{+\alpha_s}{16\pi\eps}(T_a)_i^{i'}(T_a)^j_{j'}
  (\bar q_{i'}\gamma^\mu\gamma^\alpha\otimes
  \gamma_\alpha\gamma_\mu q^{j'}),\nonumber\\[7pt]
\Delta(q^i\otimes C\bar q_j^T)
  &=&\frac{+\alpha_s}{16\pi\eps}(T_a)^i_{i'}(T_a)_j^{j'}
  (\gamma^\alpha\gamma^\mu q^{i'}\otimes
  \gamma_\alpha\gamma_\mu C\bar q_{j'}^T),\nonumber\\[7pt]
\Delta(\bar q_i\otimes C\bar q_j^T)
  &=&\frac{-\alpha_s}{16\pi\eps}(T_a)_i^{i'}(T_a)_j^{j'}
  (\bar q_{i'}\gamma^\mu\gamma^\alpha\otimes
  \gamma_\alpha\gamma_\mu C\bar q_{j'}^T),\nonumber\\[7pt]
\Delta(q^{iT}C\otimes C\bar q_j^T)
  &=&\frac{+\alpha_s}{16\pi\eps}(T_a)^i_{i'}(T_a)_j^{j'}
  (q^{i'T}C\gamma^\mu\gamma^\alpha\otimes
  \gamma_\alpha\gamma_\mu C\bar q_{j'}^T).\qquad
\end{eqnarray}
When applied to the current under question, these changes will constitute the
counterterms. Take for instance the simplified scalar current in
Eq.~(\ref{eq:skalaarvool}) and use
\begin{equation}
(T_a)^i_{i'}(T_a)^j_{j'}=\frac12\left(\delta^i_{j'}\delta^j_{j'}
  -\frac1{N_c}\delta^i_{i'}\delta^j_{j'}\right).
\end{equation}
One obtains six counterterm contributions to the currents corresponding to the
six possibilities in which the quark/antiquark lines in Fig.~\ref{tetcor} can
be connected. Starting from the top we enumerate the quark/antiquark lines
by~1 to~4.
\begin{eqnarray}
\Delta J_{(12)}&=&\frac{-\alpha_s}{2\pi\eps}\bigg[
  (u^{iT}C\gamma_5d^j)(\bar u_j\gamma_5C\bar d_i^T)
  -\frac1{N_c}(u^{iT}C\gamma_5d^j)(\bar u_i\gamma_5C\bar d_j^T)\bigg],
  \nonumber\\
\Delta J_{(13)}&=&\frac{\alpha_sC_F}{16\pi\eps}
  (u^{iT}C\gamma_5d^j)(\bar u_i\gamma_5C\bar d_j^T)
  -\frac{\alpha_sC_F}{16\pi\eps}
  (u^{iT}C\sigma^{\alpha\beta}\gamma_5d^j)
  (\bar u_i\sigma_{\alpha\beta}\gamma_5C\bar d_j^T),\nonumber\\
\Delta J_{(14)}&=&\frac{\alpha_s}{32\pi\eps}\bigg[
  (u^{iT}C\gamma_5d^j)(\bar u_j\gamma_5C\bar d_i^T)
  -\frac1{N_c}(u^{iT}C\gamma_5d^j)(\bar u_i\gamma_5C\bar d_j^T)\bigg]
  +\strut\nonumber\\&&\strut
  +\frac{\alpha_s}{32\pi\eps}\bigg[(u^{iT}C\sigma^{\alpha\beta}\gamma_5d^j)
  (\bar u_j\gamma_5\sigma_{\alpha\beta}C\bar d_i^T)
  -\frac1{N_c}(u^{iT}C\sigma^{\alpha\beta}\gamma_5d^j)
  (\bar u_i\gamma_5\sigma_{\alpha\beta}C\bar d_j^T)\bigg],\nonumber\\
\Delta J_{(23)}&=&\frac{\alpha_s}{32\pi\eps}\bigg[
  (u^{iT}C\gamma_5d^j)(\bar u_j\gamma_5C\bar d_i^T)
  -\frac1{N_c}(u^{iT}C\gamma_5d^j)(\bar u_i\gamma_5C\bar d_j^T)\bigg]
  +\strut\nonumber\\&&\strut
  +\frac{\alpha_s}{32\pi\eps}\bigg[(u^{iT}C\gamma_5\sigma^{\alpha\beta}d^j)
  (\bar u_j\sigma_{\alpha\beta}\gamma_5C\bar d_i^T)
  -\frac1{N_c}(u^{iT}C\gamma_5\sigma^{\alpha\beta}d^j)
  (\bar u_i\sigma_{\alpha\beta}\gamma_5C\bar d_j^T)\bigg],\nonumber\\
\Delta J_{(24)}&=&\frac{\alpha_sC_F}{16\pi\eps}
  (u^{iT}C\gamma_5d^j)(\bar u_i\gamma_5C\bar d_j^T)
  -\frac{\alpha_sC_F}{16\pi\eps}
  (u^{iT}C\gamma_5\sigma^{\alpha\beta}d^j)
  (\bar u_i\gamma_5\sigma_{\alpha\beta}C\bar d_j^T),\nonumber\\
\Delta J_{(34)}&=&\frac{-\alpha_s}{2\pi\eps}\bigg[
  (u^{iT}C\gamma_5d^j)(\bar u_j\gamma_5C\bar d_i^T)
  -\frac1{N_c}(u^{iT}C\gamma_5d^j)(\bar u_i\gamma_5C\bar d_j^T)\bigg].\qquad
\end{eqnarray}
The notation is such that ($mn$) stands for the diagram where the gluon is
exchanged between line $m$ and line $n$ ($m,n=1,\ldots,4$). Obviously, new
current structures have appeared which are not present at LO. Counterterms
for the currents lead to counterterms for the correlators. The whole
procedure has been automated using {\tt MATHEMATICA}.

\section{Results}
The procedure to obtain the final results for the spectral density will be
explained in detail for the colour- and flavour-antisymmetric scalar tetraquark
current $J_{S_3}$ in Eq.~(\ref{flasym}). The current consists of two parts,
\begin{equation}
J_{S_3}(x)=J_{Sa}(x)-J_{Sb}(x)
\end{equation}
where
\begin{eqnarray}
J_{Sa}(x)&=&(u^{iT}(x)C\gamma_5d^j(x))(\bar u_i(x)\gamma_5C\bar d_j^T(x)),
  \label{eq:currenttsa}\\
J_{Sb}(x)&=&(u^{iT}(x)C\gamma_5d^j(x))(\bar u_j(x)\gamma_5C\bar d_i^T(x)).
  \label{eq:currenttsb}
\end{eqnarray}
Accordingly, for the correlator one obtains four contributions.

\subsection{Diagonal contributions}
The two LO diagonal contributions $\Pi_{Saa}(x)$ and $\Pi_{Sbb}(x)$ are given
by a product of two factors $4x^2(S_0(-x^2))^2$ from the two Dirac traces and
two factors $N_c$ from the two (distinct) colour traces, resulting in
$\Pi_{Saa}^0(x)=16x^4N_c^2(S_0(-x^2))^4=\Pi_{Sbb}^0(x)$. The NLO diagonal
contribution consists of two parts. Each of the four propagator corrections is
given by the product of the same two Dirac trace factors $4x^2(S_0(-x^2))^2$
and two colour factors $N_c$. Including the general factor in
Eq.~(\ref{eq:prop_parand}),
\begin{equation}
-\frac{\alpha_sC_F}{4\pi}(-\mu_x^2x^2)^\eps\pfrac1\eps,
\end{equation}
the propagator correction reads
\begin{equation}
\Pi_{Saa}^1(x)=64N_c^2x^4\left(S_0(-x^2)\right)^4
  \left\{-\frac{\alpha_sC_F}{4\pi}(-\mu_x^2x^2)^\eps\pfrac1\eps\right\}
  =\Pi_{Sbb}^1(x).
\end{equation}
For the dipropagator corrections to the diagonal contribution one has to keep
in mind that gluon insertions are allowed only within the same colour trace.
In these two cases (in the case of $\Pi_{Saa}(x)$ for the insertions $(13)$ and
$(24)$, cf.\ Appendix~D) the colour factor is given by $\frac12N_c(N_c^2-1)$.
In calculating the Dirac traces one has to distinguish between the three parts
in Eq.~(\ref{eq:diprop_parand}), i.e.\ the parts $\slx\otimes\slx$,
$\gamma^\mu\otimes\gamma_\mu$, and $x_\mu x_\nu\gamma^{[\alpha}\gamma^\beta
\gamma^{\mu]}\otimes\gamma_{[\alpha}\gamma_\beta\gamma^{\nu]}$. Because the
gluon insertions connect different Dirac traces, one obtains the result
$16x^4(S_0(-x^2))^4$ for the first two parts and $0$ for the third one. The
dipropagator correction, therefore, reads ($N_c^2-1=2N_cC_F$)
\begin{eqnarray}
\Pi_{Saa}^2(x)&=&16N_c(N_c^2-1)x^4\left(S_0(-x^2)\right)^4
  \left\{\frac{\alpha_s}{4\pi}(-\mu_x^2x^2)^\eps
  \left[\left(\frac1\eps+\frac{11}2\right)
  +\left(\frac1\eps+\frac12\right)\right]\right\}\nonumber\\
  &=&64N_c^2C_Fx^4\left(S_0(-x^2)\right)^4\left\{\frac{\alpha_s}{4\pi}
  (-\mu_x^2x^2)^\eps\left(\frac1\eps+3\right)\right\}\ =\ \Pi_{Sbb}^2(x) .
\end{eqnarray}
Altogether one obtains
\begin{eqnarray}
\Pi_{Saa}(x)&=&16N_c^2x^4\left(S_0(-x^2)\right)^4\left\{1
  -\frac{\alpha_sC_F}\pi\eps(-\mu_x^2x^2)^\eps\pfrac1\eps
  +\frac{\alpha_sC_F}\pi(-\mu_x^2x^2)^\eps
  \left(\frac1\eps+3\right)\right\}\nonumber\\
  &=&16N_c^2x^4\left(S_0(-x^2)\right)^4\left\{1
  +3\frac{\alpha_sC_F}\pi(-\mu_x^2x^2)^\eps\right\}
  \ =\ \Pi_{Sbb}(x).
\end{eqnarray}
The diagonal contributions are finite and the counterterms are zero,
$\Pi_{Saa}^c(x)=\Pi_{Sbb}^{c}(x)=0$. Therefore, the correlator is given
by
\begin{equation}
\Pi_{Saa}(x)=16N_c^2x^4\left(S_0(-x^2)\right)^4\left\{1+\frac{4\alpha_s}\pi
  \right\}=\Pi_{Sbb}(x)
\end{equation}
and the spectral density reads
\begin{equation}
\rho_{Saa}(s)=\frac{s^4}{20(4\pi)^6}\left\{1+\frac{4\alpha_s}\pi\right\}
  =\rho_{Sbb}(s).
\end{equation}

\subsection{Nondiagonal contributions}
For the nondiagonal contributions $\Pi_{Sab}(x)$ and $\Pi_{Sba}(x)$, there
are still the same two distinct Dirac traces but only one single-colour trace
running through all lines. Because of this, the LO nondiagonal contributions
consist of a colour factor $-N_c$ and again two Dirac factors
$4x^2(S_0(-x^2))^2$, resulting in
$\Pi_{Sab}^0(x)=16N_cx^4(S_0(-x^2))^4=\Pi_{Sba}^0(x)$. Again, the same
factors occur also for the four propagator corrections. Including the general
factor from Eq.~(\ref{eq:prop_parand}) one obtains
\begin{equation}\label{Sab1}
\Pi_{Sab}^1(x)=64N_cx^4\left(S_0(-x^2)\right)^4
  \left\{-\frac{\alpha_sC_F}{4\pi}(-\mu_x^2x^2)^\eps\pfrac1\eps\right\}
  =\Pi_{Sba}^1(x).
\end{equation}
For the dipropagator corrections, each of the six gluon insertions leads to
a colour factor $\frac12(N_c^2-1)=N_cC_F$. The Dirac traces, however,
depend on whether the gluon insertions are within the same Dirac trace or not.
For the insertions $(12)$ and $(34)$ the contributions to the three parts of
Eq.~(\ref{eq:diprop_parand}) are given by $-16x^4(S_0(-x^2))^4$,
$-16Dx^4(S_0(-x^2))^4$, and $-16(D-1)(D-2)x^4(S_0(-x^2))^4$, respectively. For
the other four insertions we obtain again $16x^4(S_0(-x^2))^4$,
$16x^4(S_0(-x^2))^4$, and $0$. Including the general factor from
Eq.~(\ref{eq:diprop_parand}) the dipropagator corrections read
\begin{eqnarray}\label{Sab2}
\Pi_{Sab}^2(x)&=&64N_cx^4\left(S_0(-x^2)\right)^4\left\{
  -\frac{\alpha_sC_F}{4\pi}(-\mu_x^2x^2)^\eps\left[\left(\frac4\eps+1\right)
  -2\left(\frac1\eps+3\right)\right]\right\}\nonumber\\
  &=&64N_cx^4\left(S_0(-x^2)\right)^4\left\{
  -\frac{\alpha_sC_F}{4\pi}(-\mu_x^2x^2)^\eps\left(\frac2\eps-5\right)\right\}
  \ =\ \Pi_{Sba}^2(x).
\end{eqnarray}
Altogether one obtains
\begin{eqnarray}
\Pi_{Sab}(x)&=&16N_cx^4\left(S_0(-x^2)\right)^4\left\{1
  -4\frac{\alpha_s}\pi(\mu_x^2x^2)^\eps\left(\frac1\eps-\frac53\right)
  \right\}\nonumber\\
  &=&16N_cx^4\left(S_0(-x^2)\right)^4\left\{1-4\frac{\alpha_s}\pi
  \left(\frac1\eps-\frac53+\ln(\mu_x^2x^2)\right)\right\}\ =\ \Pi_{Sba}(x).
\end{eqnarray}
The nondiagonal contributions are singular. Following the considerations in
Sec.~3 the corresponding $x$-space counterterms can be obtained in the same
way as the first-order propagator and dipropagator corrections in
Eqs.~(\ref{Sab1}--\ref{Sab2}). One keeps only the singular contribution
which contributes with the opposite sign to those in
Eqs.~(\ref{Sab1}--\ref{Sab2}). The counterterms, therefore, are given by
\begin{equation}
\Pi_{Sab}^c(x)=64N_cx^4\left(S_0(-x^2)\right)^4\frac{\alpha_s}\pi\pfrac1\eps
  =\Pi_{Sba}^c(x).
\end{equation}
The renormalized nondiagonal correlator contributions read
\begin{equation}
\Pi_{Sab}^r(x)=16N_cx^4\left(S_0(-x^2)\right)^4\left\{1-4\frac{\alpha_s}\pi
  \left(\ln(\mu_x^2x^2)-\frac53\right)\right\}=\Pi_{Sba}^r(x).
\end{equation}
When summing up all four contributions one obtains
\begin{eqnarray}
\Pi_{S_3S_3}(x)&=&\Pi_{Saa}-\Pi_{Sab}-\Pi_{Sba}+\Pi_{Sbb}
  \ =\ 2\Pi_{Saa}-2\Pi_{Sab}\nonumber\\[3pt]
  &=&192x^4\left(S_0(-x^2)\right)^4\left\{1+2\frac{\alpha_s}\pi
  (\mu_x^2x^2)^\eps\left(\frac1\eps+\frac43\right)\right\}\nonumber\\
  &=&192x^4\left(S_0(-x^2)\right)^4\left\{1+2\frac{\alpha_s}\pi
  \left(\frac1\eps+\frac43+\ln(\mu_x^2x^2)\right)\right\}.
\end{eqnarray}
The resulting counterterm is thus given by
\begin{equation}
\Pi_{S_3S_3}^c(x)=-384x^4\left(S_0(-x^2)\right)^4
  \frac{\alpha_s}\pi\pfrac1\eps,
\end{equation}
leading to a renormalized correlator of the form
\begin{equation}
\Pi_{S_3S_3}^r(x)=384x^4\left(S_0(-x^2)\right)^4\left\{1+2\frac{\alpha_s}\pi
  \left(\frac43+\ln(\mu_x^2x^2)\right)\right\}.
\end{equation}
The corresponding spectral density (cf.\ Appendix~E) reads
\begin{equation}
\rho_{S_3S_3}(s)=\frac{s^4}{15(4\pi)^6}\left\{1+\frac{\alpha_s}\pi
  \left(\frac{57}5+2\ln\pfrac{\mu_\msbar^2}s\right)\right\}.
\end{equation}

\subsection{Results for diagonal and nondiagonal spectral functions}
The same procedure works for all currents. Let us first list all ten diagonal
spectral functions corresponding to the five flavour-antisymmetric and five
flavour-symmetric currents. One obtains
\begin{eqnarray}
\rho^\sigma_{S_3S_3}=\ \rho^\sigma_{P_3P_3}&=&\frac{s^4}{15(4\pi)^6}
  \left\{1+\frac{\alpha_s}\pi\left(\frac{57}5
  +2\ln\pfrac{\mu_\msbar^2}s\right)\right\},
  \nonumber\\
\rho^\sigma_{V_3V_3}\ =\ \rho^\sigma_{A_3A_3}&=&\frac{4s^4}{15(4\pi)^6}
  \left\{1+\frac{\alpha_s}\pi\left(\frac{67}{10}
  +\ln\pfrac{\mu_\msbar^2}s\right)\right\},
  \nonumber\\
\rho^\sigma_{T_3T_3}&=&\frac{8s^4}{5(4\pi)^6}
  \left\{1+\frac{\alpha_s}\pi\left(\frac{82}{15}
  +\frac23\ln\pfrac{\mu_\msbar^2}s\right)\right\},\nonumber\\
\rho^\sigma_{S_6S_6}=\ \rho^\sigma_{P_6P_6}&=&\frac{2s^4}{15(4\pi)^6}
  \left\{1+\frac{\alpha_s}\pi\left(\frac3{10}
  -\ln\pfrac{\mu_\msbar^2}s\right)\right\},
  \nonumber\\
\rho^\sigma_{V_6V_6}\ =\ \rho^\sigma_{A_6A_6}&=&\frac{8s^4}{15(4\pi)^6}
  \left\{1+\frac{\alpha_s}\pi\left(\frac{55}4
  +\frac52\ln\pfrac{\mu_\msbar^2}s\right)\right\},
  \nonumber\\
\rho^\sigma_{T_6T_6}&=&\frac{16s^4}{5(4\pi)^6}
  \left\{1+\frac{\alpha_s}\pi\left(\frac{211}{15}
  +\frac{11}3\ln\pfrac{\mu_\msbar^2}s\right)\right\}.
\end{eqnarray}
The LO contributions for the flavour-antisymmetric spectral densities are in
agreement with the results of Ref.~\cite{Chen:2007xr}. The LO contributions
for the flavour-symmetric spectral densities amount to twice the corresponding
antisymmetric contributions as can be understood from the colour manipulations
described in Appendix~D. For example, one has
$\rho^\sigma_{S_6S_6}({\rm LO})=2\rho^\sigma_{S_3S_3}({\rm LO})$, etc.

When considering spectral functions corresponding to mixed currents as e.g.\
in Eq.~(\ref{mixed}) of Sec.1.1 one also needs nondiagonal spectral functions. 
For the example~(\ref{mixed}) one needs the nondiagonal spectral functions
\begin{equation}
\rho^\sigma_{A_6V_3}=\rho^\sigma_{V_3A_6}=\rho^\sigma_{A_3V_6}
=\rho^\sigma_{V_6A_3}=\frac{8s^4}{15(4\pi)^6}
  \left\{-\frac{\alpha_s}\pi\left(\frac{48}5+3\ln\pfrac{\mu_\msbar^2}s\right)
  \right\}.
\end{equation}
Finally, we list all nonvanishing nondiagonal spectral functions,
\begin{eqnarray}
\rho^\sigma_{V_3A_3}=\ -\rho^\sigma_{S_3T_3}&=&\frac{4s^4}{5(4\pi)^6}
  \pfrac{\alpha_s}{6\pi},\nonumber\\
\rho^\sigma_{V_6A_6}=\ -\rho^\sigma_{S_6T_6}&=&\frac{4s^4}{5(4\pi)^6}
  \pfrac{5\alpha_s}{6\pi},\nonumber\\
\rho^\sigma_{S_3T_6}=\ \rho^\sigma_{S_6T_3}
  \ =\ -\rho^\sigma_{V_3A_6}\ =\ -\rho^\sigma_{V_6A_3}
&=&\frac{4s^4}{5(4\pi)^6}\pfrac{\alpha_s}\pi
  \left(\frac{16}5+\ln\pfrac{\mu_\msbar^2}s\right),\nonumber\\
\rho^\sigma_{T_3P_6}=\ \rho^\sigma_{T_6P_3}
&=&\frac{-4s^4}{5(4\pi)^6}\pfrac{\alpha_s}\pi
  \left(\frac{37}{10}+\ln\pfrac{\mu_\msbar^2}s\right).
\end{eqnarray}

\section{Summary and conclusion}
We have obtained analytical results for the NLO perturbative contributions
to the light tetraquark correlation and spectral functions.
The results have been obtained by prudently hopping back and forth
between $p$ space and $x$ space making use of the modular approach introduced
in Refs.~\cite{Groote:1998wy,Groote:1999cx} in terms of propagator and
dipropagator insertions into the correlation functions. We have checked on the
gauge invariance of our results. At the same time we have also checked on the
gauge invariance of previous results on the textbook case of zero-mass meson
correlators~\cite{Reinders:1984sr,Narison:1989aq,Muta:1998vi}, on baryon 
correlators~\cite{Ovchinnikov:1988zx,Ovchinnikov:1991mu} and on pentaquark
correlators~\cite{Groote:2006sy,Groote:2011my}.

We have found that the NLO perturbative corrections to the correlators are 
large. However, since the current correlators are dominated by the
nonperturbative contributions, the NLO corrections have little impact on the
analysis of the corresponding sum rules. In particular, referring to
Appendix~F, we find that the NLO corrections to the perturbative term affect
the results of the sum rule analysis for the ground-state energy by $+0.065\%$
which is within the error of the Borel sum rule analysis by Chen {\it et
al.}~\cite{Chen:2007xr}. However, as pointed out in Sec.~1, the error on the 
condensate contributions used in Ref.~\cite{Chen:2007xr} may have been vastly
underestimated. A different possibility to harness the large condensate
contributions of Ref.~\cite{Chen:2007xr} would be to analyze the spectral
functions in terms of finite-energy sum
rules~\cite{Pivovarov:1991rh,Groote:1997cn,Groote:1997kh} in which the large
higher-twist condensate contributions are reduced or even removed. 

In calculating the NLO contributions to the pentaquark correlators we have
set the light-quark masses to zero. We do not expect quark-mass effects
as e.g.\ the strange-current-quark mass to be important for the NLO corrections
at the scale of the light scalar mesons. This may be different in the sum rule
analysis of heavy tetraquark states where the nonperturbative contributions can
be expected to be smaller. We hope to return to the problem of calculating the
perturbative corrections to heavy tetraquark current correlators in the future,
using again a modular approach.

A possible further project would be to calculate the radiative corrections to
the LO dibaryon (``exiquark'') correlator discussed in
Ref.~\cite{Larin:1985yt}.

\subsection*{Acknowledgments}
This work was supported by the Estonian Research Council under grant
No.~IUT2-27, and by the Estonian Science Foundation under grant No.~8769.
We would like to thank A.~Grozin and A.A.~Pivovarov for useful discussions.
S.G.\ acknowledges the support by the Mainz Institute of Theoretical Physics
(MITP). 

\begin{appendix}

\section{The scalarity of the diquark current}
\setcounter{equation}{0}\def\theequation{A\arabic{equation}}
In Table~\ref{Flo:bilineaarsed} we list the five bilinear quark--quark
currents that are being used to construct the tetraquark currents. $C$ is 
the charge conjugation
matrix given by $C=i\gamma^2\gamma^0$, and the index $T$ stands for
transposition. Because we are dealing with diquarks (or antidiquarks), the
labeling of the currents in terms of their parity properties differs from 
the familiar labeling of bilinear quark--antiquark fields. In the following
we show that the diquark current
\begin{equation}\label{eq:diqvool}
J_S(x)=q^{T}(x)C\gamma_5q(x)
\end{equation}
is a scalar current. This can be demonstrated similarly to the textbook example
of proving the scalarity of the quark-antiquark current $q(x)\bar q(x)$. 

Using the Lorentz transformation property
$x^\mu\rightarrow x^{\prime\mu}=\Lambda^\mu{}_\nu x^\nu =(\Lambda x)^\mu$
one can write
\begin{equation}
q(x)\to q'(x')=q'(\Lambda x)=U(\Lambda)q(x),
\end{equation}
We then use the fact that the metric tensor and the Dirac equation are 
invariant under Lorentz transformations,
\begin{equation}
\Lambda^Tg\Lambda=g,\qquad
U^{-1}(\Lambda)\gamma^\mu U(\Lambda)=\Lambda^\mu{}_\nu\gamma^\nu.
\end{equation}
Next we expand
$U(\Lambda)=\bbbone+\frac i4\epsilon_{\alpha\beta}h^{\alpha\beta}+O(h^2)$
where $\epsilon_{\alpha\beta}$ is the two-dimensional Levi--Civita symbol
and where $h^{\alpha\beta}$ is an antisymmetric infinitesimal quantity.

Let us briefly return to the textbook example of the quark--antiquark current. 
Using the properties $\gamma^0\gamma^0=\bbbone$,
$\gamma^0\gamma^{\mu\dagger}\gamma^0=\gamma^\mu$ and
$\gamma^0\sigma^{\mu\nu\dagger}\gamma^0=\sigma^{\mu\nu}$ one can show
that $\gamma^0U^\dagger\gamma^0=U^{-1}$ and, therefore,
\begin{equation}
\bar q'q'=q^\dagger U^\dagger\gamma^0Uq=\bar q U^{-1}Uq=\bar qq,
\end{equation}
showing that $\bar q(x)q(x)$ transforms as a scalar.

\begin{table}
\begin{center}
\begin{tabular}{|c|c|c|}
\hline 
\noalign{\vskip\doublerulesep}
scalar&$S$&$q^TC\gamma_5q$
 \tabularnewline[\doublerulesep]\hline\noalign{\vskip\doublerulesep}
vector&$V$&$q^TC\gamma^\mu\gamma_5q$
\tabularnewline[\doublerulesep]\hline\noalign{\vskip\doublerulesep}
tensor&$T$&$q^TC\sigma^{\mu\nu}q$
 \tabularnewline[\doublerulesep]\hline\noalign{\vskip\doublerulesep}
axial vector&$A$&$q^TC\gamma^\mu q$
 \tabularnewline[\doublerulesep]\hline\noalign{\vskip\doublerulesep}
pseudoscalar&$P$&$q^TCq$
 \tabularnewline[\doublerulesep]\hline
\end{tabular}\end{center}
\caption{\label{Flo:bilineaarsed}Bilinear diquark Dirac fields}
\end{table}

In the case of the diquark current in Eq.~(\ref{eq:diqvool}) we define
$A=C\gamma_5=i\gamma^3\gamma^1$. Contrary to the charge conjugation matrix
$C=i\gamma^2\gamma^0$ where $CC = -\bbbone$, $C\gamma^{\mu T}C=\gamma^\mu$ and
$C\sigma^{\mu\nu T}C=\sigma^{\mu\nu}$, one obtains $AA=\bbbone$,
$A\gamma^{\mu T}A=\gamma^\mu$ and $A\sigma^{\mu\nu T}A=-\sigma^{\mu\nu}$.
Therefore, for the matrix $U$ one obtains $AU^TA=U^{-1}$, and one can conclude
that
\begin{equation}
q^{\prime T}C\gamma_5q'=q^{\prime T}Aq'=q^TU^TAUq
  =q^TAU^{-1}Uq=q^TC\gamma_5q,
\end{equation}
showing that $q^TC\gamma_5q$ transforms as a scalar. One could have 
anticipated this result by intuitive reasoning from the fact that two quark 
fields have a relative positive parity while a quark and antiquark field have
relative negative quality.

\section{Fierz transformations}
\setcounter{equation}{0}\def\theequation{B\arabic{equation}}
We shall present the generalized Fierz transformation \cite{Fierz:1937} in
terms of the Takahashi bracket notation~\cite{Takahashi:1986} which reads 
\begin{equation}\label{fierz}
(\Gamma^A)[\Gamma^B]=\frac1{4^2}
  \Tr(\Gamma^A\Gamma_C\Gamma^B\Gamma_D)(\Gamma^D][\Gamma^C).
\end{equation}
The summation runs over the indices $C$ and $D$. The bracket notation is best
explained by writing out the corresponding Dirac indices,
\begin{equation}
(\Gamma^A)[\Gamma^B]=\Gamma^A_{\alpha\beta}\Gamma^B_{\gamma\delta},\qquad
(\Gamma^A][\Gamma^B)=\Gamma^A_{\alpha\delta}\Gamma^B_{\gamma\beta}.
\end{equation}
Let us introduce a set of five Dirac strings
\begin{equation}\label{gamma}
\Gamma^A\in\{\bbbone,\gamma^\mu,\sigma^{\mu\nu},\gamma^\mu\gamma_5,\gamma_5\},
\end{equation}
where $\sigma^{\mu\nu}=\frac i2[\gamma^\mu,\gamma^{\nu}]$. The dual base to
this set has the elements $\Gamma_A$ where
$\Tr(\Gamma_A\Gamma^B)=4\delta_A^B$.\footnote{In the case of multi-indices $A$
and $B$ as for instance for $\Gamma_A=\sigma_{\mu\nu}$,
$\Gamma^B=\sigma^{\rho\sigma}$ one has to use the convention
$\delta_{\mu\nu}^{\rho\sigma}=\delta_\mu^\rho\delta_\nu^\sigma
-\delta_\mu^\sigma\delta_\nu^\rho$.} We obtain (note the different order in
the axial term and the factor $1/2$ for the tensor part)
\begin{equation}
\Gamma_A\in\{\bbbone,\gamma_\mu,\Frac12\sigma_{\mu\nu},
  \gamma_5\gamma_\mu,\gamma_5\}.
\end{equation}
Next we define a set of five contracted outer products of the Dirac strings 
in Eq.~(\ref{gamma}),
\begin{equation}
(S,V,T,A,P)
  =\Big((1)[1],(\gamma^\mu)[\gamma_\mu],(\sigma^{\mu\nu})[\sigma_{\mu\nu}],
  (\gamma^\mu\gamma_5)[\gamma_\mu\gamma_5],(\gamma_5)[\gamma_5]\Big)
\end{equation}
and their Fierz-reordered counterparts
\begin{equation}
(\tilde S,\tilde V,\tilde T,\tilde A,\tilde P)=
  \Big((1][1),(\gamma^\mu][\gamma_\mu),(\sigma^{\mu\nu}][\sigma_{\mu\nu}),
  (\gamma^\mu\gamma_5][\gamma_\mu\gamma_5),(\gamma_5][\gamma_5)\Big).
\end{equation}
The two sets are related by the Fierz transformation matrix. The elements
of the matrix can be calculated with the help of Eq.~(\ref{fierz}). This will
be illustrated for the last row of the Fierz transformation matrix whose 
coefficients can be calculated from the trace $16(\gamma_5)[\gamma_5]
=\Tr(\gamma_5\Gamma_A\gamma_5\Gamma_B)(\Gamma^A][\Gamma^B)$. Only the 
diagonal terms contribute to the trace. One obtains 
\begin{eqnarray}
\Gamma^A=\bbbone,\ \Gamma^B=\bbbone&:&\Tr(\gamma_5\gamma_5)=4,\nonumber\\ 
\Gamma^A=\gamma^\mu,\ \Gamma^B=\gamma^\nu&:&
  \Tr(\gamma_5\gamma_\mu\gamma_5\gamma_\nu)=-4g_{\mu\nu},\nonumber\\
\Gamma^A=\sigma^{\mu\nu},\ \Gamma^B=\sigma^{\rho\sigma}&:&
  \Tr(\gamma_5\Frac12\sigma_{\mu\nu}\gamma_5\Frac12\sigma_{\rho\sigma})
  =g_{\mu\rho}g_{\nu\sigma}-g_{\mu\sigma}g_{\nu\rho},\nonumber\\
\Gamma^A=\gamma_5\gamma^\mu,\ \Gamma^B=\gamma_5\gamma^\nu&:&
  \Tr(\gamma_5\gamma_5\gamma_\mu\gamma_5\gamma_5\gamma_\nu)=4g_{\mu\nu},
  \nonumber\\
\Gamma^A=\gamma_5,\ \Gamma^B=\gamma_5&:&
  \Tr(\gamma_5\gamma_5\gamma_5\gamma_5)=4,
\end{eqnarray}
which implies
\begin{equation}
16(\gamma_5)[\gamma_5]=4(\bbbone][\bbbone)-4(\gamma^\mu][\gamma_\mu)
  +2(\sigma^{\mu\nu}][\sigma_{\mu\nu})
  +4(\gamma^\mu\gamma_5][\gamma_\mu\gamma_5)+4(\gamma_5][\gamma_5).
\end{equation}
The other elements of the Fierz crossing matrix can be calculated accordingly.

The transformation between these two sets is given by the Fierz matrix which
reads
\begin{equation}\label{fierz1}
\left(\begin{array}{c}S\\V\\T\\A\\P\end{array}\right)
  =\frac14\left(\begin{array}{ccccc}1&1&1/2&-1&1\\4&-2&0&-2&-4\\12&0&-2&0&12\\
  -4&-2&0&-2&4\\1&-1&1/2&1&1\end{array}\right)\left(\begin{array}{c}
  \tilde S\\\tilde V\\\tilde T\\\tilde A\\\tilde P\end{array}\right).
 \end{equation}

Using the Fierz matrix~(\ref{fierz1}) we can determine the relation
between the diquark--antidiquark interpolating currents
\begin{eqnarray}
({\cal S},{\cal V},{\cal T},{\cal A},{\cal P})
  &=&\Big((u^{iT}C\gamma_5d^j)(\bar u_i\gamma_5C\bar d_j^T),
  (u^{iT}C\gamma^\mu\gamma_5d^j)(\bar u_i\gamma_\mu\gamma_5C\bar d_j^T),\\&&
  (u^{iT}C\sigma^{\mu\nu}d^j)(\bar u_i\sigma_{\mu\nu}C\bar d_j^T),
  (u^{iT}C\gamma^\mu d^j)(\bar u_i\gamma_\mu C\bar d_j^T),
  (u^{iT}C d^j)(\bar u_i C\bar d_j^T)\Big)\nonumber
\end{eqnarray}
and the meson--meson-type interpolating currents
\begin{eqnarray}
(\tilde{\cal S},\tilde{\cal V},\tilde{\cal T},\tilde{\cal A},\tilde{\cal P})
  &=&\Big((\bar d_ju^i)(\bar u_id^j),
  (\bar d_j\gamma^\mu u^i)(\bar u_i\gamma_\mu d^j),
  (\bar d_j\sigma^{\mu\nu}u^i)(\bar u_i\sigma_{\mu\nu}d^j),\nonumber\\&&
  (\bar d_j\gamma^\mu\gamma_5u^i)(\bar u_i\gamma_\mu\gamma_5 d^j),
  (\bar d_j\gamma_5u^i)(\bar u_i\gamma_5d^j)\Big).
\end{eqnarray}
Note the relabeling in going from the sets of Dirac strings $(S,V,T,A,P)$
and $(\tilde S,\tilde V,\tilde T,\tilde A,\tilde P)$ to the sets of current 
products $({\cal S},{\cal V},{\cal T},{\cal A},{\cal P})$ and 
$(\tilde{\cal S},\tilde{\cal V},\tilde{\cal T},\tilde{\cal A},\tilde{\cal P})$.

By making use of the properties $CC=-\bbbone$, $C^{T}=-C$,
$C\gamma_5C=-\gamma_5= -\gamma_5^T$, $C\gamma^\mu C=\gamma^{\mu T}$,
$C\gamma_5\gamma^\mu C=-(\gamma_5\gamma^\mu)^T$ and
$C\sigma^{\mu\nu}C=\sigma^{\mu\nu T}$ one obtains e.g.\ for the first row 
of the transformation matrix
\begin{eqnarray}
\lefteqn{(u^{iT}C\gamma_5d^j)(\bar u_i\gamma_5C\bar d_j^T)
  =-\frac14(\bar d_ju^i)(\bar u_id^j)
  -\frac14(\bar d_j\gamma^\mu u^i)(\bar u_i\gamma_\mu d^j)}\nonumber\\&&
  +\frac18(\bar d_j\sigma^{\mu\nu}u^i)(\bar u_i\sigma_{\mu\nu}d^j)
  -\frac14(\bar d_j\gamma^\mu\gamma_5u^i)(\bar u_i\gamma_\mu\gamma_5d^j)
  -\frac14(\bar d_j\gamma_5u^i)(\bar u_i\gamma_5d^j).
\end{eqnarray}
A similar exercise allows one to calculate the remaining coefficients.

The above two sets of interpolating currents are thus related by 
\begin{equation}\label{fierz2}
\left(\begin{array}{c}{\cal S}\\{\cal V}\\{\cal T}\\{\cal A}\\{\cal P}
  \end{array}\right)
  =\frac14\left(\begin{array}{ccccc}-1&-1&1/2&-1&-1\\4&-2&0&2&-4\\
  -12&0&-2&0&-12\\-4&-2&0&2&4\\-1&1&1/2&1&-1\end{array}\right)
  \left(\begin{array}{c}\tilde{\cal S}\\\tilde{\cal V}\\\tilde{\cal T}\\
  \tilde{\cal A}\\\tilde{\cal P}\end{array}\right).
\end{equation}
Note that the Fermion fields are commuted four times in this transformation
such that the overall sign resulting from the Fermi statistics is positive.

\section{Fourier transform in dimensional regularization}
\setcounter{equation}{0}\def\theequation{C\arabic{equation}}
In order to calculate the $D$-dimensional integral of a Lorentz scalar, we
need to know, among others, the $(D-1)$-dimensional angular integral of a 
Lorentz scalar. In the Euclidean domain one has
($\kappa=(\tilde k^{2})^{1/2}$)
\begin{equation}
\int f(\tilde k^2)d^D\tilde k
  =\int d\Omega\int_0^\infty f(\kappa^2)\kappa^{D-1}d\kappa
  =\frac{2\pi^{D/2}}{\Gamma(D/2)}\int_0^\infty f(\kappa^2)\kappa^{D-1}d\kappa,
\end{equation}
where
\begin{equation}
\int d\Omega=\int_0^\pi\frac{2\pi^{(D-1)/2}}{\Gamma((D-1)/2)}
  \sin^{D-2}\theta\,d\theta=\frac{2\pi^{D/2}}{\Gamma(D/2)}
\end{equation}
and $\Gamma(x)$ is Euler's gamma function. Using Euler's beta function
\begin{equation}
B(x,y):=\int_{0}^1t^{x-1}(1-t)^{y-1}dt=\frac{\Gamma(x)\Gamma(y)}{\Gamma(x+y)}
\end{equation}
and the definite integrals
\begin{equation}
\int_{-1}^{+1}e^{ixt}(1-t^2)^{\lambda-1/2}dt
  =\pfrac2x^\lambda\sqrt\pi J_\lambda(x),\quad
\int_0^\infty x^\mu J_\lambda(x)dx
  =2^\mu\frac{\Gamma((1+\lambda+\mu)/2)}{\Gamma((1+\lambda-\mu)/2)}
\end{equation}
($J_{\lambda}(x)$ is Bessel's function), one can show that
\begin{equation}
\int\frac{d^D\tilde k}{(2\pi)^D}(\tilde k^2)^{-\alpha}e^{i\tilde k\tilde x}
  =\frac{\Gamma(D/2-\alpha)}{4\pi^{D/2}\Gamma(\alpha)}
  \pfrac{\tilde x^2}4^{\alpha-D/2}.
\end{equation}
With $k_0=i\tilde k_0$ and $x_0=i\tilde x_0$ one gets back to the Minkowskian
domain where $k^2=-\tilde k^2$ and $x^2=-\tilde x^2$. The result is 
\begin{equation}\label{eq:konfruum1}
\int\dDk(-k^2)^{-\alpha}e^{-ikx}
  =\frac{i\Gamma(D/2-\alpha)}{(4\pi)^{D/2}\Gamma(\alpha)}
  \left(-\frac{x^2}4\right)^{\alpha-D/2}.
\end{equation}
By applying the partial derivatives $\partial_\mu=\partial/\partial x^\mu$,
$\partial_\nu=\partial/\partial x^\nu$ on both sides of 
Eq.~(\ref{eq:konfruum1}) one obtains 
\begin{equation}\label{eq:konfruum2}
\int\dDk(-k^2)^{-\alpha}e^{-ikx}k_\mu
  =-\frac{\Gamma(D/2-\alpha+1)}{2(4\pi)^{D/2}\Gamma(\alpha)}
\left(-\frac{x^2}4\right)^{\alpha-D/2-1}x_\mu
\end{equation}
and
\begin{eqnarray}\label{eq:konfruum3}
\lefteqn{\int\dDk(-k^2)^{-\alpha}e^{-ikx}k_{\mu}k_{\nu}
  \ =\ \frac{i\Gamma(D/2-\alpha+1)}{8(4\pi)^{D/2}\Gamma(\alpha)}}\nonumber\\&&
  \times\left(-\frac{x^2}4\right)^{\alpha-D/2-2}
  \left[2(\alpha-D/2-1)x_\mu x_\nu+x^2g_{\mu\nu}\right]
\end{eqnarray}
and, finally,
\begin{eqnarray}\label{eq:konfruum4}
\lefteqn{\int\dDk(-k^2)^{-\alpha}e^{-ikx}(Ak_\mu k_\nu+Bk^2g_{\mu\nu})
  \ =\ \frac{i\Gamma(D/2-\alpha+1)}{8(4\pi)^{D/2}\Gamma(\alpha)}}\nonumber\\&&
  \times\left(-\frac{x^2}4\right)^{\alpha-D/2-2}\bigg[2A
  (\alpha-D/2-1)x_\mu x_\nu+\Big(A+2B(\alpha-1)\Big)x^2g_{\mu\nu}\bigg].
\end{eqnarray}

\section{Gauge independence} 
\setcounter{equation}{0}\def\theequation{D\arabic{equation}}
In this appendix we present results on the propagator and dipropagator
corrections calculated in the $R_{\xi}$ gauge where the gluon propagator
reads
\begin{equation}
D_{\alpha\beta}(k)=\frac{i}{k^2}\left(-g_{\alpha\beta}
  +(1-\xi)\frac{k_{\alpha}k_{\beta}}{k^2}\right).
\end{equation}
The momentum-dependent piece proportional to
$(1-\xi)k_{\alpha}k_{\beta}/{k^2}$ will be referred to as the scalar part
of the gluon propagator. We shall show that the scalar contribution vanishes
in the sum of the propagator and dipropagator insertions into the correlators
of colour-neutral currents (mesons, baryons and tetraquarks). We believe that
the gauge independence of the radiative corrections to the correlators have
never been demonstrated before. The gauge independence can be shown without
specifying the Dirac structure of the currents. Our results on the tetraquark
correlators are thus gauge independent for any of the currents discussed in
the main text. The gauge independence of the NLO correlators also serves as a
strong check on our calculation.

The calculation of the scalar contribution to the one-loop propagator
correction does not provide any new difficulties compared to the metric
contribution. For the dipropagator correction the scalar part of the gluon
propagator superficially increases the rank of the tensor two-loop integrals
to six. However, by a prudent cancellation of numerator and denominator
factors one can reduce the rank to two as in the contribution of the metric
piece. As a check on our two-loop calculation we did an alternative
calculation involving sixth-rank tensor integrals which we solved using the 
Passarino--Veltman method. We found agreement. We mention that all necessary
calculations have been checked by computer.

We shall demonstrate the gauge invariance of the NLO radiative corrections for
meson, baryon and tetraquark correlators. In the general $R_\xi$ gauge, the
propagator correction reads ($C_F=(N_c^2-1)/(2N_c)$)
\begin{equation}\label{Pi1x}
S_1^\xi(x)=S_0(-x^2)\left\{1-\frac{\alpha_sC_F}{4\pi}(-\mu_x^2x^2)^\eps
  \Big(1-(1-\xi)\Big)\left(\frac1\eps+O(\eps)\right)+O(\alpha_s^2)\right\}
  x_\mu\gamma^\mu
\end{equation}
while the dipropagator correction is given by
\begin{eqnarray}\label{dipropcorc}
\lefteqn{S_2^\xi(x)\ =\ \left(S_0(-x^2)\right)^2\Bigg[x_\mu x_\nu
  (\gamma^\mu\otimes\gamma^\nu)-(T_a\otimes T_a)\frac{\alpha_s}{4\pi}
  (-\mu_x^2x^2)^\eps\times}\nonumber\\&&\times
  \Bigg\{\left(\left(\frac1\eps+\frac{11}2-(1-\xi)\frac2\eps+O(\eps)\right)
  x_\mu x_\nu+\left(\frac1\eps+\frac12+O(\eps)\right)x^2g_{\mu\nu}\right)
  (\gamma^\mu\otimes\gamma^\nu)+\strut\nonumber\\&&\strut
  +\left(\frac1{2\eps}+\frac14+O(\eps)\right)x_\mu x_\nu
  (\gamma^{[\mu}\gamma^\alpha\gamma^{\beta]}\otimes
  \gamma^{[\nu}\gamma_\alpha\gamma_{\beta]})\Bigg\}+O(\alpha_s^2)\Bigg].
\end{eqnarray}
We have written the results in a form where the contribution of the
scalar piece of the gluon propagator proportional to $(1-\xi)$ can be clearly
identified. In the Landau (or unitary) gauge $\xi=1$ one has the familiar
result that the propagator correction vanishes.

Note the essential fact that both gauge dependencies occur as pure
singularities in the contributions $x_\mu\gamma^\mu$ (propagator) and
$x_\mu\gamma^\mu\otimes x_\nu\gamma^\nu$ (dipropagator) and that the gauge
dependence of the dipropagator correction amounts to twice the gauge 
dependence of the propagator correction. Note also that both gauge-dependent
corrections to the propagator and the dipropagator are UV singular.

In the following we concentrate on the gauge-dependent scalar contribution
proportional to $(1-\xi)$. Using the notation $\slx=x_\mu\gamma^\mu$ one has
\begin{eqnarray}
\Delta S_1^\xi(x)&=&S_0(-x^2)\frac{\alpha_sC_F}{4\pi}(-\mu_x^2x^2)^\eps
  (1-\xi)\left(\frac1\eps\right)\slx,\nonumber\\
\Delta S_2^\xi(x)&=&\left(S_0(-x^2)\right)^2(T_a\otimes T_a)
  \frac{\alpha_s}{4\pi}(-\mu_x^2x^2)^\eps(1-\xi)\left(\frac2\eps\right)
  \slx\otimes\slx.
\end{eqnarray}
It is important to realize that both gauge-dependent corrections have the
spatial structure of the respective LO term. Note also that both corrections
are UV singular.

We start our discussion with the meson case. The demonstration of gauge
invariance is made simple in $x$ space. The gauge-dependent part of the
NLO propagator correction to a meson correlator augmented by the free
propagator reads
$\Delta S_M^\xi(x)=\Delta S_1^\xi(x)\otimes S_1^0(-x)
+S_1^0(x)\otimes\Delta S_1^\xi(-x)$, or
\begin{equation}
\Delta S_{M1}^\xi(x)=-2N_c\left(S_0(-x^2)\right)^2\frac{\alpha_sC_F}{4\pi}
  (-\mu_x^2x^2)^\eps(1-\xi)\left(\frac1\eps\right)\slx\otimes\slx.
\end{equation}
Note that there is an extra minus sign from the antiquark propagator. Also 
one needs the colour factor $\delta_i^j\delta^i_j=N_c$. The factor of two 
results from the fact that the propagator correction can be inserted into the
quark or antiquark line.

For the dipropagator correction one requires the colour factor 
\begin{equation}
\delta_{i'}^{j'}(T_a)^{i'}_i(T_a)_{j'}^j\delta^i_j
  =\frac12\delta_{i'}^{j'}\left(\delta^{i'}_{j'}\delta_i^j
  -\frac1{N_c}\delta^{i'}_i\delta_{j'}^j\right)\delta^i_j
  =\frac12(N_c^2-1)=N_cC_F.
\end{equation}
There is no extra minus sign since there are two antiquark lines, one each on 
either side of the quark--gluon vertex. One obtains 
\begin{equation}
\Delta S_{M2}^\xi(x)=N_cC_F\left(S_0(-x^2)\right)^2\frac{\alpha_s}{4\pi}
  (-\mu_x^2x^2)^\eps(1-\xi)\left(\frac2\eps\right)\slx\otimes\slx.
\end{equation}
Obviously, the two contributions cancel in the sum,
$\Delta S_{M1}^\xi(x)+\Delta S_{M2}^\xi(x)=0$.

For the baryon correlator the gauge-dependent part of the propagator
correction reads
\begin{equation}
\Delta S_{B1}^\xi(x)=3N_c!\left(S_0(-x^2)\right)^3\frac{\alpha_sC_F}{4\pi}
  (-\mu_x^2x^2)^\eps(1-\xi)\left(\frac1\eps\right)\slx\otimes\slx\otimes\slx.
\end{equation}
The factor $\epsilon_{ijk}\epsilon^{ijk}=N_c!$ results from the colour
contraction while the factor $3$ has to be included because of the three
quark lines into which the propagator correction can be inserted. For the
dipropagator insertion one needs the colour factor
\begin{equation}
\epsilon_{i'j'k}(T_a)^{i'}_i(T_a)^{j'}_j\epsilon^{ijk}
  =\frac12\epsilon_{i'j'k}\left(\delta^{i'}_j\delta_i^{j'}
  -\frac1{N_c}\delta^{i'}_i\delta^{j'}_j\right)\epsilon^{ijk}=-N_c!C_B,
\end{equation}
where $C_B=(N_c+1)/(2N_c)=C_F/2$. Again there are three possible dipropagator
insertions resulting in a further factor of $3$. One obtains
\begin{equation}
\Delta S_{B2}^\xi(x)=-3N_c!C_B\left(S_0(-x^2)\right)^3\frac{\alpha_s}{4\pi}
  (-\mu_x^2x^2)^\eps(1-\xi)\left(\frac2\eps\right)\slx\otimes\slx\otimes\slx.
\end{equation}
The two contributions can be seen to cancel, i.e. 
$\Delta S_{B1}^\xi(x)+\Delta S_{B2}^\xi(x)=0$.

Finally, we demonstrate the gauge parameter cancellation for the tetraquark 
correlator. We label the two quark and antiquark lines of the in-state by the
colour indices $(i,j)$ and $(k,l)$. We associate the indices
$(i,j,k,l)$ with the (first, second, third, fourth) line of the tetraquark
state (see \ Fig.~\ref{tetcor}(a)) starting at the top. The corresponding 
labeling in the out-state is $(i',j')$ and $(k',l')$ with the same sequence
in the numerical labeling. In the meson-type construction the colour-singlet
tetraquark states are given by $\delta^i_k\delta^j_l$ for the in-state and
$\delta_{i'}^{k'}\delta_{j'}^{l'}$ for the out-state. However, as discussed
in Sec.~\ref{inter} we need to separate out the antisymmetric 
{\bf $\bar 3\oplus 3$} and symmetric {\bf $6\oplus \bar 6$} colour components
of the currents. This is achieved by writing 
\begin{equation}
\label{sepin}
\delta^i_k\delta^j_l=\frac12\underbrace{\left(\delta^i_k\delta^j_l
  -\delta^j_k\delta^i_l\right)}_{{\bf\bar 3\oplus 3}}
  +\frac12\underbrace{\left(\delta^i_k\delta^j_l
  +\delta^j_k\delta^i_l\right)}_{{\bf 6\oplus\bar 6}}
\end{equation}
for the in-state and, correspondingly,
\begin{equation}
\label{sepout}
\delta_{i'}^{k'}\delta_{j'}^{l'}=
  \frac12\left(\delta_{i'}^{k'}\delta_{j'}^{l'}
  -\delta_{i'}^{l'}\delta_{j'}^{k'}\right)
  +\frac12\left(\delta_{i'}^{k'}\delta_{j'}^{l'}
  +\delta_{i'}^{l'}\delta_{j'}^{k'}\right)
\end{equation}
for the out-state.

The propagator correction can be inserted into the correlator in four ways
leading to a factor of $4$.
We thus obtain
\begin{equation}\label{propsc}
\Delta S_{T1}^\xi(x)=4C_{T1}\left(S_0(-x^2)\right)^4\frac{\alpha_sC_F}{4\pi}
  (-\mu_x^2x^2)^\eps(1-\xi)\left(\frac1\eps\right)\slx\otimes\slx\otimes\slx
  \otimes\slx,
\end{equation}
where
\begin{eqnarray}
C_{T1}(3\,\to3)&=&\tfrac12N_c(N_c-1),\qquad
C_{T1}(6\,\to6)\ =\ \tfrac12N_c(N_c+1),\\
C_{T1}(3\,\to6)&=&C_{T1}(6\,\to3)\ =\ 0.\nonumber 
\end{eqnarray}
Since we are considering also nondiagonal $(3\to6)$ and $(6\to3)$ transitions
in the main text, we list the corresponding colour factors also for the
nondiagonal cases even if they are trivially zero for the propagator 
correction. This is no longer the case for the dipropagator corrections
to be discussed next.

The dipropagator correction can be inserted into the correlator in six 
different ways. We shall label these six different possibilities by the
lines that are being connected by the gluon propagator as described in 
Sec.~3. For example, the labeling $(13)$ refers to gluon exchange between
the top and third line (from the top) as depicted in Fig.~\ref{tetcor}(c). 
In general one has
\begin{equation}
\Delta S_{T2}^\xi(x)=C_{T2}\left(S_0(-x^2)\right)^4\frac{\alpha_s}{4\pi}
  (-\mu_x^2x^2)^\eps(1-\xi)\left(\frac2\eps\right)\slx\otimes\slx\otimes\slx
  \otimes\slx,
\end{equation}
where the factor $C_{T2}$ specifies the colour factor of a given gluon 
connection including the factor $(-1)^{n_{\bar q}}$ resulting from the presence
of $n_{\bar q}$ antiquark lines in that particular transition. For example, 
the colour factor in the $(3\,\to3)$ $(12)$ contribution including the factor
$(-1)^{n_{\bar q}}$ is given by ($\Tr_c(T_aT_a)=N_cC_F$)
\begin{equation}
C_{T2}(12;3\to3)=\frac12\left(\delta^i_k\delta^j_l-\delta^j_k\delta^i_l\right)
  (-1)^{2}\Big((T_a)_i^{i'}(T_a)_j^{j'}\delta^k_{k'}\delta^l_{l'}\Big)
  \frac12\left(\delta_{i'}^{k'}\delta_{j'}^{l'}
  -\delta_{i'}^{l'}\delta_{j'}^{k'}\right)=-\frac12N_cC_F.
\end{equation}
Similarly, the colour kernel for the $(13)$ contribution is given by
$(-1)^{1}((T_a)_i^{i'}\delta_j^{j'}(T_{a})^k_{k'}\delta^l_{l'})$.

The colour factors for the different line connections and transitions are
listed in Table~\ref{tab2}. Of relevance for the present discussion is the
respective sum of the six rows in Table~\ref{tab2} which are listed in the
seventh row of Table~\ref{tab2}. From the last row of Table~\ref{tab2} one can
read off that the gauge-dependent nondiagonal $(3\to6)$ and $(6\to3)$
transitions are zero as expected. The gauge-dependent diagonal parts given
by the propagator correction Eq.~(\ref{propsc}) and the dipropagator
correction in the last row of Table~\ref{tab2} can be seen to cancel. We
mention that we have checked on the gauge cancellation also for pentaquark
current correlators investigated in Refs.~\cite{Groote:2006sy,Groote:2011my}.

It is important to realize that, in the Feynman gauge calculation discussed
in Sec.~4, one requires the colour factors in Table~\ref{tab2} for each row 
separately since their contributions carry different weights due to the
new spatial non-Born structures in the dipropagator correction 
Eq.~(\ref{eq:diprop3}).

\begin{table}
\begin{center}
\begin{tabular}{|r||c|c|c|c||c|}\hline
$C_{T2}$&$3\to3$&$3\to6$&$6\to3$&$6\to6$&sum\\\hline\hline
$(12)$&$-2$&$0$&$0$&$2$&$0$\\\hline
$(13)$&$-N_c+2$&$-N_c$&$-N_c$&$-N_c-2$&$-4N_c$\\\hline
$(14)$&$-N_c+2$&$N_c$&$N_c$&$-N_c-2$&$0$\\\hline
$(23)$&$-N_c+2$&$N_c$&$N_c$&$-N_c-2$&$0$\\\hline
$(24)$&$-N_c+2$&$-N_c$&$-N_c$&$-N_c-2$&$-4N_c$\\\hline
$(34)$&$-2$&$0$&$0$&$2$&$0$\\\hline\hline
sum&$-4(N_c-1)$&$0$&$0$&$-4(N_c+1)$&$-8N_c$\\\hline
\end{tabular}
\caption{\label{tab2}Colour factor $C_{T2}$ for the different dipropagator
insertions $(12),\ldots,(34)$ and the diagonal $3\to3$ and $6\to6$ and the
nondiagonal $3\to6$ and $6\to3$ transitions. All entries have to be multiplied
by a general factor $N_cC_F/4$. The entries contain also the sign factor
$(-1)^{n_{\bar q}}$ due to the number of antiquark lines. The last column
contains the sum of the four first columns corresponding to the colour
contraction given by Eqs.~(\ref{sepin}) and~(\ref{sepout}). In the last row we
list the sum of the six first rows.}
\end{center}
\end{table}

\section{The spectral density}
\setcounter{equation}{0}\def\theequation{E\arabic{equation}}
In this appendix we derive relations which allow us to calculate the spectral 
density directly from the correlator in $x$ space. For the scalar correlator
the transition to $p$ space is given by 
\begin{equation}
\Pi(p)=2\pi^{\lambda+1}\int_0^\infty\pfrac{px}2^{-\lambda}
  J_\lambda(px)\Pi(x)x^{2\lambda+1}dx,
\end{equation}
where $\lambda=1-\eps$ and $J_\lambda(x)$ is the first-order Bessel function. 
The arguments $x$ and $p$ are not four-vectors, but rather (in the Euclidean
domain) the lengths of the vectors, i.e. $x=(x_\mu x^\mu)^{1/2}$ and
$p=(p_\mu p^\mu)^{1/2}$.

If the correlator is a given by a simple power, $\Pi(x)=(-x^2)^{-\alpha}$,
the integral can be calculated to be
\begin{equation}
\Pi_\alpha(-p^2)=\pi^{\lambda+1}\left(-\frac{p^2}4\right)^{\alpha-\lambda-1}
  \frac{\Gamma(\lambda-\alpha+1)}{\Gamma(\alpha)}.
\end{equation}
The spectral density is the discontinuity divided by $2\pi i$, where the 
cut of the correlator lies on the positive real axis. One obtains
\begin{equation}\label{eq:rhoalpha}
\rho_\alpha(s)=\frac1{2\pi i}\Disc\Pi_\alpha(s)=\pi^{\lambda+1}
  \pfrac s4^{\alpha-\lambda-1}\frac1{\Gamma(\alpha)\Gamma(\alpha-\lambda)}.
\end{equation}
For tetraquarks the $x$-space correlator has the generic form 
\begin{equation}
f(x)=(S_0(-x^2))^4(x^2)^2\left\{A+\frac{\alpha_s}{\pi}
  \left(-\mu_x^2x^2\right)^\eps B\right\},
\end{equation}
where $A$ includes both the LO term and the counterterm while $B$ includes
only the NLO term. Keeping in mind that
$S_0(-x^2)=i\Gamma(2-\eps)/2(-\pi x^2)^{2-\eps}$, one can apply
Eq.~(\ref{eq:rhoalpha}) to obtain
\begin{eqnarray}
\lefteqn{\rho_f(x)\ =\ \pi^{2-\eps}\pfrac{i\Gamma(2-\eps)}{2\pi^{2-\eps}}^4
  \frac{(s/4)^{4-3\eps}}{\Gamma(6-4\eps)\Gamma(5-3\eps)}}\nonumber\\&&
  \times\left\{A+\frac{\alpha_s}{\pi}\pfrac{\mu_\msbar^2}s^\eps
  e^{2\gamma_E\eps}B\frac{\Gamma(6-4\eps)\Gamma(5-3\eps)}{\Gamma(6-5\eps)
  \Gamma(5-4\eps)}\right\},
\end{eqnarray}
where we have used $4\mu_x^2\approx\mu_\msbar^2e^{2\gamma_E}$. The ratio of
gamma functions can be expanded by using
$\Gamma(a+\eps)=\Gamma(a)(1+\eps\psi(a)+O(\eps^2))$,
where $\psi(z)=\Gamma'(z)/\Gamma(z)$ is the digamma function. One obtains
\begin{equation}
e^{2\gamma_E\eps}
  \frac{\Gamma(6-4\eps)\Gamma(5-3\eps)}{\Gamma(6-5\eps)\Gamma(5-4\eps)}
  =1-\frac{131}{30}\eps+O(\eps^2)
\end{equation}
and, therefore,
\begin{equation}
\rho_f(s)=\frac{\Gamma(2-\eps)^4s^{4-3\eps}}{4\pi^{6-3\eps}
  \Gamma(6-4\eps)\Gamma(5-4\eps)}\left\{A+\frac{\alpha_s}{\pi}
  \pfrac{\mu_\msbar^2}s^\eps B\left(1-\frac{131}{30}\eps\right)\right\}.
\end{equation}
By separating the finite and singular parts of $A$ and $B$ one has 
\begin{equation}
A=A_0+\frac{\alpha_s}\pi\left(\frac{C_0}\eps+C_1\right),\qquad
B=\frac{B_0}\eps+B_1.
\end{equation} 
With $B_0+C_0=0$ one obtains
\begin{eqnarray}\label{rhof}
\rho_f(s)&=&\frac{\Gamma(2-\eps)^4s^{4-3\eps}}{4\pi^{6-3\eps}
  \Gamma(6-4\eps)\Gamma(5-4\eps)}\left\{A_0+\frac{\alpha_s}{\pi}
  \left(B_1+C_1-\frac{130}{30}B_0+B_0\ln\pfrac{\mu_\msbar^2}s\right)\right\}
  \nonumber\\
  &=&\frac{s^4}{(4\pi)^65!4!}\left\{A_0+\frac{\alpha_s}{\pi}
  \left(B_1+C_1-\frac{130}{30}B_0+B_0\ln\pfrac{\mu_\msbar^2}s\right)\right\}.
\end{eqnarray}
Because the spectral function $\rho_f(s)$ is nonsingular, we have set 
$\eps=0$ (i.e.\ $D=4$) in the second line of Eq.~(\ref{rhof}).

\section{QCD sum rule analysis}
\setcounter{equation}{0}\def\theequation{F\arabic{equation}}
In this appendix we provide a brief review of the sum rule method using the
Borel transformation. The starting expression for the analysis is the sum rule
\begin{equation}\label{F1}
\frac{F_X^2}{E_X^2-p^2}=\int_0^{E_c^2}\frac{\rho(s)ds}{s-p^2},
\end{equation}
where $E_X$ is the ground-state energy and $F_X^2$ is the residue of the pole
at $p^2=E_X^2$. The beginning of the continuous spectrum is denoted by $E_c$.
The convergence of the sum rule can be improved by performing a Borel
transformation on both sides of Eq.~(\ref{F1}), leading to
\begin{equation}
F_X^2e^{-E_X^2/E_B^2}=\int_0^{E_c^2}\rho(s)e^{-s/E_B^2}ds,
\end{equation}
where $E_B$ is the Borel energy. For the sum rule analysis one has to search 
for an energy window in which the dependence on the artificial Borel parameter
$E_B$ is small. One expands the spectral density as a power series in $s$ and
replaces the powers $(s/E_B^2)^k/k!$ by ($x_C=E_C^2/E_B^2$)
\begin{equation}
f_k(x_C)=\int_0^{x_C}\frac{x^{\prime k}}{k!}e^{-x}dx
  =1-e^{-x_C}\sum_{m=0}^k\frac{x_C^m}{m!}.
\end{equation}
If there are logarithmic contributions, one has to replace
$\ln(s/E_B^2)(s/E_B^2)^k/k!$ by
\begin{eqnarray}
f_k^\ell(x_C)&=&\int_0^{x_C}\ln x\frac{x^k}{k!}e^{-x}dx\nonumber\\
  &=&\ln x_C(f_k(x_C)-1)-\gamma_E+\Ei(-x_C)+\sum_{m=0}^{k-1}\frac{f_m(x_C)}m,
\end{eqnarray}
where
\begin{equation}
\gamma_E=-\int_0^\infty\ln x\,e^{-x}dx
\end{equation}
is Euler's constant and where $\Ei(x_C)$ is an exponential integral given by
\begin{equation}
\Ei(x_C)=\int_{-x_C}^\infty\frac{e^{-x}}xdx.
\end{equation}
Writing the operator product expansion of the spectral density in the form
\begin{equation}
\rho(s)=A_4s^4\left\{1+\frac{\alpha_s}\pi
  \left(\alpha+\beta\ln\pfrac{\mu_\msbar^2}s\right)\right\}
  +A_3s^3+A_2s^2+A_1s+A_0,
\end{equation}
one obtains
\begin{eqnarray}\label{borel}
\lefteqn{F_X^2e^{-E_X^2/E_B^2}\ =\ \int_0^{E_C^2}\rho(s)e^{-s/E_B^2}ds
\ =}\nonumber\\
  &=&4!A_4E_B^{10}\left\{f_4(x_C)+\frac{\alpha_s}\pi\left(
  \alpha f_4(x_C)+\beta\left(\ln\pfrac{\mu_\msbar^2}{E_B^2}
  f_4(x_C)-f_4^\ell(x_C)\right)\right)\right\}\nonumber\\[7pt]&&\strut
  +3!A_3E_B^8f_3(x_C)+2!A_2E_B^6f_2(x_C)
  +1!A_1E_B^4f_1(x_C)+0!A_0E_B^2f_0(x_C).\qquad
\end{eqnarray}
The ground-state energy $E_X$ can be determined by calculating the derivative
of Eq.~(\ref{borel}) with respect to $-1/E_B^2$ and then dividing the
derivative by Eq.~(\ref{borel}),
\begin{equation}
E_X^2=\frac{\displaystyle\int_0^{E_C^2}s\rho(s)e^{-s/E_B^2}ds}{\displaystyle
  \int_0^{E_C^2}\rho(s)e^{-s/E_B^2}ds}.
\end{equation}
For the derivative one obtains
\begin{eqnarray}
\lefteqn{F_X^2E_X^2e^{-E_X^2/E_B^2}
  \ =\ \int_0^{E_C^2}s\rho(s)e^{-s/E_B^2}ds\ =}\nonumber\\
  &=&4!A_4E_B^{10}\Bigg(5E_B^2\left\{f_4(x_C)
  +\frac{\alpha_s}\pi\left(\alpha f_4(x_C)
  +\beta\left(\ln\pfrac{\mu_\msbar^2}{E_B^2}f_4(x_C)
  -f_4^\ell(x_C)\right)\right)\right\}\nonumber\\&&\strut
  -E_C^2\left\{f'_4(x_C)+\frac{\alpha_s}\pi\left(\alpha f'_4(x_C)
  +\beta\left(\frac1{x_C}f_4(x_C)+\ln\pfrac{\mu_\msbar^2}{E_B^2}f'_4(x_C)
  -f^{\ell\prime}_4(x_C)\right)\right)\right\}\Bigg)\nonumber\\[7pt]&&\strut
  +3!E_B^8A_3\left(4E_B^2f_3(x_C)-E_C^2f'_3(x_C)\right)
  +2!E_B^6A_2\left(3E_B^2f_2(x_C)-E_C^2f'_2(x_C)\right)\nonumber\\[7pt]&&\strut
  +1!E_B^4A_1\left(2E_B^2f_1(x_C)-E_C^2f'_1(x_C)\right)
  +0!E_B^2A_0\left(E_B^2f_0(x_C)-E_C^2f'_0(x_C)\right).\qquad
\end{eqnarray}
The analysis is performed with the same parameters for the Borel window as
in Ref.~\cite{Chen:2007xr}. The addition of radiative corrections changes the
result of the sum rule analysis for the ground-state energy by $+0.065\%$
which is within the error of the sum rule analysis.


\end{appendix}

\end{document}